\documentclass[notitlepage,nofootinbib,preprintnumbers,amssymb,superscriptaddress]{revtex4-2}
\usepackage{graphicx}
\usepackage{dcolumn}
\usepackage{bm}
\usepackage{comment}
\usepackage{amsfonts,amssymb,mathtools,graphicx,color,bm}
\definecolor{ultramarine}{rgb}{0.07, 0.04, 0.56}
\definecolor{cadmiumgreen}{rgb}{0.0, 0.42, 0.24}
\definecolor{indigo(dye)}{rgb}{0.0, 0.25, 0.42}
\usepackage[linktocpage=true]{hyperref}
\hypersetup{
colorlinks=true,
citecolor=ultramarine,
linkcolor=cadmiumgreen,
urlcolor=indigo(dye),
}

\usepackage{autobreak}

\newcommand{\de}{\, \mathrm{d}}

\makeatletter

\renewcommand*{\p@subsection}{}

\renewcommand*{\p@subsubsection}{}
\makeatother

\numberwithin{equation}{section}

\begin{document}
\begin{flushleft} {\footnotesize Dedicated to the memorial issue in IJMPA of V.~A.~Rubakov} \end{flushleft}
\
\begin{flushright} {\footnotesize YITP-25-05} \end{flushright}

\affiliation{School of Physics Science and Engineering, Tongji University, Shanghai 200092, China}
\affiliation{Institute for Advanced Study of Tongji University, Shanghai 200092, China}

\author{Misao Sasaki}
\email{misao.sasaki@ipmu.jp}
\affiliation{Kavli Institute for the Physics and Mathematics of the Universe (WPI), UTIAS, The University of Tokyo, Chiba 277-8583, Japan}
\affiliation{Asia Pacific Center for Theoretical Physics, Pohang 37673, Korea}
\affiliation{Center for Gravitational Physics and Quantum Information, Yukawa Institute for Theoretical Physics, Kyoto University, Kyoto 606-8502, Japan}
\affiliation{Leung Center for Cosmology and Particle Astrophysics, National Taiwan University, Taipei 10617, Taiwan}

\author{Vicharit Yingcharoenrat}
\email{vicharit.y@chula.ac.th}
\affiliation{Kavli Institute for the Physics and Mathematics of the Universe (WPI), UTIAS, The University of Tokyo, Chiba 277-8583, Japan}
\affiliation{High Energy Physics Research Unit, Department of Physics, Faculty of Science, Chulalongkorn University, Pathumwan, Bangkok 10330, Thailand}

\author{Ying-li Zhang}
\email{yingli@tongji.edu.cn}
\affiliation{School of Physics Science and Engineering, Tongji University, Shanghai 200092, China}
\affiliation{Institute for Advanced Study of Tongji University, Shanghai 200092, China}
\affiliation{Kavli Institute for the Physics and Mathematics of the Universe (WPI), UTIAS, The University of Tokyo, Chiba 277-8583, Japan}
\affiliation{
    Institute of Theoretical Physics,
    Chinese Academy of Sciences,
    Beijing 100190, China
}
\affiliation{Center for Gravitation and Cosmology, Yangzhou University, Yangzhou 225009, China}

\title{Singular instantons with finite action}

\begin{abstract}
Recently, it was shown that in the absence of gravity there exist non-$O(4)$-symmetric instanton solutions with finite action beyond Coleman's instantons. In this paper, focusing on the false-vacuum decay in a single scalar field in flat Euclidean space, we provide a general discussion on 
$O(4)$-symmetric instantons that are singular at the true-vacuum bubble. We find that, for the action to remain finite without introducing a UV cutoff, the potential must be unbounded from below, thereby evading Coleman's theorem. We then consider two explicit examples of such instantons and perturbatively analyze the dynamics of small deformations around them. We find that one of them does not allow regular deformations, which indicates that the $O(4)$ symmetric solution still gives the minimum action, while the other one is found to allow regular deformations that cost no additional action at second order in perturbation. The latter example opens up the possibility of the existence of non-linear non-$O(4)$-symmetric solutions with lower action if we allow singular instantons with finite action.

\end{abstract}

\maketitle

\section{Introduction}
False-vacuum decay is a non-perturbative phenomenon in quantum physics with significant implications for both particle physics and cosmology. 
In the absence of gravitational effects, such a tunneling process is described by a classical solution to the equations of motion in Euclidean space, called instanton solutions. These solutions provide a physical framework for understanding the transition from a metastable (false) vacuum to a stable (true) vacuum ~\cite{Coleman:1977py,Callan:1977pt}. 
In a single scalar-field model the decay rate $\Gamma$ per unit time per unit volume ($\mathcal{V}$) in the semiclassical approximation is given by 
\begin{align}\label{eq:decay_rate}
    \frac{\Gamma}{\mathcal{V}}  = A\,e^{-B[\Phi_{\rm cl}]} \;,
\end{align}
where $A$ is the prefactor that contains $\hbar$ corrections and $B[\Phi_{\rm cl}]$ is the bounce action evaluated on a classical solution. 
We define the bounce action via the on-shell action $S[\Phi_{\rm cl}]$ as
\begin{align}\label{eq:bounce_action}
    B[\Phi_{\rm cl}] \equiv S[\Phi_{\rm cl}] - S[\Phi_{\rm FV}] \;,
\end{align}
with $\Phi_{\rm FV}$ being the location of the false vacuum.  
Note that a classical solution is subjected to appropriate boundary conditions, as we will discuss below.
In the present paper we are only interested in the leading order in $\hbar$, which is the exponent of the decay rate (\ref{eq:decay_rate}).\footnote{The prefactor $A$, in principle, can be computed using the functional determinant of the second variation of the action evaluated on classical solutions. This is subdominant compared to the exponent.}

In \cite{Coleman:1977th} it was proven in a rigorous manner that in the absence of gravity the $O(4)$-symmetric solutions yield the minimal bounce action (the most probable tunneling process) under certain conditions on the potential.
Note that the proof was done for a canonical real scalar field with potential $V(\Phi)$ in a $d$-dimensional Euclidean space.
Specifically, the potential $V(\Phi)$ has to be admissible, i.e., it is differentiable and continuous for all $\Phi$, and it is bounded from below.
More precisely, the condition for the potential to be bounded from below reads $V \geq a|\Phi|^\alpha + b|\Phi|^\beta$ for all $\Phi$ where $a$, $b$, $\alpha$ and $\beta$ are positive numbers satisfying $\alpha < \beta < 2d/(d-2)$.
The fact that one requires the potential to be bounded from below is because the potential is guaranteed to have a true vacuum within a finite range of $\Phi$.
These assumptions of the proof \cite{Coleman:1977th} therefore correspond to the following boundary conditions for $\Phi$,
\begin{align}\label{eq:con_coleman}
\Phi(\rho)\big|_{\rho \rightarrow \infty} = \Phi_{\rm FV} \;,  \quad \frac{\mathrm{d}\Phi(\rho)}{\mathrm{d}\rho} \bigg|_{\rho \rightarrow 0}= 0 \;,
\end{align}
where $\rho$ denotes the radial coordinate in the 4-dimensional Euclidean space. 
The conditions above ensure that in Euclidean space the solution starts at rest at the true vacuum $(\rho \to 0)$ and reaches the false vacuum as $\rho \to \infty$.
We see that the second condition in (\ref{eq:con_coleman}) guarantees that the solution $\Phi(\rho)$ is regular at the center of the true-vacuum bubble. 
We call Euclidean solutions that satisfy the condition (\ref{eq:con_coleman}) the Coleman instantons.

It was recently pointed out in \cite{Mukhanov:2021rpp,Mukhanov:2021kat,Espinosa:2019hbm} that the Coleman instantons do not exist for a class of (unbounded) potentials in an arbitrary spacetime dimension. 
Instead, they found a class of new solutions that do not satisfy the boundary conditions (\ref{eq:con_coleman}), see \cite{Mukhanov:2020pau,Mukhanov:2020wim,Mukhanov:2021ggy} for new instantons\footnote{In this case, the UV cutoff was introduced to regularize the solution, so that the action is finite.} and \cite{Espinosa:2019hbm,Espinosa:2021qeo} for pseudo bounces. 
For further developments in the study of the thick-wall approximation for unbounded potentials, see \cite{Mukhanov:2022abt,Matteini:2024xvg}. 
More recently, in \cite{Sasaki:2024vul} we demonstrated that in the presence of unbounded exponential potential there exist non-trivial and regular deformations around a singular and $O(4)$-symmetric instanton solution, whose action is the same as the $O(4)$-symmetric ones. 
Although these are infinitesimal deformations, the existence of them indicates that there may exist non-linear non-$O(4)$-symmetric instanton solutions (beyond Coleman's instantons).
The situation explained above becomes even more complicated when gravity is taken into account. Here we do not include gravity.  

In this paper, as a complementary version to \cite{Sasaki:2024vul}, we discuss in more detail all possible ways to obtain singular instanton solutions with finite action.\footnote{This is somewhat similar to the case of Hawking-Turok instanton \cite{Hawking:1998bn} where gravity is included. In this case, the non-trivial geometry makes the action finite, even though the instanton is singular.} 
Again, these singular solutions do not satisfy the second condition of (\ref{eq:con_coleman}), i.e., one instead starts with infinite velocity in Euclidean space ($V \to -V$).
We then derive a general form of the potential, that gives rise to singular instantons with finite action. 
As we will see, this potential is unbounded from below and violates one of the assumptions of Coleman's theorem.
Moreover, we explicitly perform a detailed analysis on the dynamics of classical deformations around singular instantons up to second order in two concrete examples: (1) a cubic potential and (2) a piecewise quadratic potential. These two potentials are smoothly connected with the unbounded exponential potential.

The rest of this paper is organized as follows.
Our setup is given in section~\ref{sec:setup}.
In section~\ref{sec:sin_instanton_deform},
we derive a general form of the potential, which gives rise to singular instanton solutions with finite action. 
To make the analysis complete, we also discuss the possibility of a solution with a cusp at the center of the true vacuum bubble.
In section~\ref{sec:deform_instan}, we analyze the dynamics of anisotropic deformations around the $O(4)$-symmetric solutions up to secound order. 
In section~\ref{sec:con_example}, we consider two specific examples, one if the potential with its the false vacuum side given by a cubic function and the other by a piecewise quadratic function. Then, we explicitly demonstrate the non-existence of deformed instanton solutions in the cubic case, and the existence of them in the piece-wise quadratic case for a specific choice of the potential parameters.
We conclude the paper in section~\ref{sec:conclusion}.

\section{Setup}\label{sec:setup}
Let us start with a real scalar field model with the Euclidean action,
\begin{align}\label{eq:action_Eu}
S = \int \mathrm{d}^4x \sqrt{g} \bigg[\frac{1}{2}\,g^{\mu\nu} \partial_\mu \Phi(x) \partial_\nu \Phi(x) + V(\Phi) \bigg] \;,
\end{align}
where $V(\Phi)$ is the potential. 
In this paper we do not include gravity and fix the metric to be the 4-dimensional flat Euclidean metric, 
\begin{align}\label{eq:flat_metric}
\mathrm{d}s^2 = \mathrm{d}\tau^2 + \mathrm{d}\vec{x}^2 = \mathrm{d}\rho^2 + \rho^2(\mathrm{d}\theta^2 + \sin^2\theta\,\mathrm{d}\Omega_2^2) \;, 
\end{align}
where $\tau$ is the Euclidean time, $\rho \equiv \sqrt{\tau^2 + \vec{x}^2}$ and $\mathrm{d}\Omega_2^2 \equiv \mathrm{d}\chi^2 + \sin^2 \chi\,\mathrm{d}\phi^2$.
Notice that although the metric after the second equality in \eqref{eq:flat_metric} is expressed in the manifestly $O(4)$ symmetric form, the model may admit solutions which are not $O(4)$-symmetric. It may also admit solutions which are singluar at the origin ($\rho=0$).

From the action (\ref{eq:action_Eu}), the equation of motion (EoM) of $\Phi$ reads 
\begin{align}
\Box \Phi - \frac{\mathrm{d}V(\Phi)}{\mathrm{d}\Phi} = 0 \;,
\label{eq:EOM_phi}
\end{align}
where $\Box \equiv g^{\mu\nu}\partial_\mu \partial_\nu$. The EoM above in general describes dynamics of instanton solution with prescribed boundary conditions: one at the origin of the bubble ($\rho = 0$) or true vacuum and the other at infinity ($\rho \rightarrow \infty$) or false vacuum. 
The tunneling rate in the semiclassical limit is given by Eq.~(\ref{eq:decay_rate}). 
Here, we are interested the computation of the bounce action~(\ref{eq:bounce_action}).

As explained in the introduction, the Coleman instanton is assumed to satisfy the boundary conditions (\ref{eq:con_coleman}), the second of which ensures its regular behavior at the center of the bubble. 
Note that the regularity condition indeed plays a crucial role in the proof of \cite{Coleman:1977th}.
In the next section, instead of imposing the second condition, we allow instantons to be singular at the origin, or equivalently the true vacuum is located infinitely far away in field space, provided that the action remains finite.
Then, we consider anisotropic solutions perturbatively around the singular $O(4)$-symmetric instanton, and show that such perturbations with non-vanishing action do not exist. Namely, either such an anisotropic solution never exists or an anisotropic deformation does not cost additional action if such a solution exists.
\section{Singular instanton and its deformation}\label{sec:sin_instanton_deform}
In this section we will analyze a general condition on the potential, under which the action (\ref{eq:action_Eu}) evaluated on the singular instanton is finite, and we will study a possibility of having a small anisotropic deformation around such instanton solutions. 

\subsection{General form of the potential}\label{sec:gen_potential}
Let us consider the $O(4)$-symmetric instanton: $\Phi = \bar{\Phi}(\rho)$. Under the Euclidean metric \eqref{eq:flat_metric}, EoM~(\ref{eq:EOM_phi}) takes the form
\begin{align}
\bar{\Phi}'' + \frac{3}{\rho} \bar{\Phi}' - \frac{\mathrm{d}V}{\mathrm{d}\Phi}\bigg|_{\Phi = \bar{\Phi}} = 0\;, \label{eq:EOM_phi_zeroth}
\end{align}
where prime is a derivative with respect to $\rho$. The corresponding Euclidean action \eqref{eq:action_Eu} reduces to 
\begin{align}\label{eq:action_zero}
S_0 =  2\pi^2 \int_0^\infty \mathrm{d}\rho~\rho^3~\bigg[\frac{1}{2}\bar{\Phi}'^2 + V(\bar{\Phi}) \bigg] \;,
\end{align}
where the factor $2\pi^2$ was obtained from the integrals over the angular variables.
Notice before that as $\rho \rightarrow \infty$ we have $\bar{\Phi} \rightarrow \Phi_{\rm FV}$ and $V(\bar{\Phi}) \rightarrow V(\Phi_{\rm FV}) = const$, which seems to give rise to an infinite action since the measure goes as $\rho^3$. 
However, this is not the case due to the fact that the decay rate is actually computed from the bounce action, see Eq.~(\ref{eq:bounce_action}).
We thus see that the convergence of the action is guaranteed in the limit $\rho \rightarrow \infty$.

Let us now discuss the limit $\rho$ goes to zero (the center of the bubble). 
Here, as mentioned before, we are interested in the case where the solution as well as its derivatives diverge as $\rho \rightarrow 0$.
With this assumption, in general we see that if $\Phi'^2_0(\rho)$ blows up faster than $1/\rho^3$ when $\rho \rightarrow 0$, the action (\ref{eq:action_zero}) inevitably goes to infinity. 
Therefore, the requirement that the action is finite yields the conditions on the behaviors of $\bar{\Phi}'(\rho)$ and $\bar{\Phi}(\rho)$ as $\rho \to 0$.

A possible case to realize this argument is $\bar{\Phi}'(\rho\to 0)\propto1/\rho$, which corresponds to a logarithmically singular behavior of $\Phi(\rho)$. 
Hence, now we investigate the condition to realize the solution $\bar{\Phi}(\rho\to0)=-M\log(\rho/\rho_\star)$, where $M>0$ is a constant and $\rho_\star$ is a pivot scale.\footnote{We will, later on, comment on other types of singular behavior of $\bar{\Phi}'$ such as power-law and exponential.} 
Since this solution approaches $+\infty$ as $\rho \rightarrow 0$, we call it ``singular instanton". Inserting $\bar{\Phi}$ into
Eq.~(\ref{eq:EOM_phi_zeroth}) we obtain
\begin{align}\label{eq:con_dV_sing}
    \lim_{\rho \rightarrow 0} \frac{\mathrm{d}V}{\mathrm{d}\Phi}\bigg|_{\Phi = \bar{\Phi}} = - \frac{2 M}{\rho^2} \;,
\end{align}
from which we can deduce that close to the center of the bubble the potential takes the form,
\begin{align}\label{eq:potential_gen_con}
    V(\Phi) = -\Lambda^4\exp\bigg(\frac{2\Phi}{M}\bigg) \;,
\end{align}
with $\Lambda^4 \equiv (M/\rho_\star)^2$. 
Notice that the form (\ref{eq:potential_gen_con}) is generic, as we have only assumed that $\bar{\Phi}(\rho)$ satisfies the EoM and diverges logarithmically in the limit $\rho \rightarrow 0$.
It is evident that the potential (\ref{eq:potential_gen_con}) is unbounded from below when $\Phi\to+\infty$. Therefore, it violates one of the assumptions of the Coleman's theorem \cite{Coleman:1977th}. 
Indeed this allows us to have other types of instanton solutions for which the value of the corresponding Euclidean action may be the same or even lower than that of the Coleman's instatons.
Additionally, it is straightforward to verify that with the potential (\ref{eq:potential_gen_con}) the integrand of the integral (\ref{eq:action_zero}) goes as $-\rho\,M^2/2$, which is vanishing as $\rho \rightarrow 0$. 
Notice that in our case one is not required to introduce a UV-cutoff as in \cite{Mukhanov:2020pau,Mukhanov:2020wim,Mukhanov:2021ggy} since our integral over $\rho$ is finite. 
Therefore, we obtain a singular $O(4)$-symmetric instanton with finite action. 

Let us now comment on the regularization of the potential $V(\Phi)$ as $\Phi \to \infty$.
We see that from the above discussion our potential is unbounded from below. 
Of course, we can always regularize the potential at very large $\Phi$, which corresponds to a small radius $\rho_0$, to obtain a regularized instanton.
However, we find that, even in the limit $\rho_0 \to 0$, the action remains finite.
Therefore, the resulting action is independent of the regularization procedure.

Before closing this section, let us comment on other kinds of singular behavior of $\bar{\Phi}(\rho)$. 
Suppose that in the limit $\rho \rightarrow 0$ the solution behaves as 
\begin{align}\label{eq:other_sing_phi0}
    \bar{\Phi}(\rho) = M \bigg(\frac{\rho_\star}{\rho}\bigg)^n \;, \quad \text{or} \quad \bar{\Phi}(\rho) = M \exp\bigg(\frac{\rho_\star}{\rho}\bigg) \;,
\end{align}
where $n$ is a positive integer.
Following the same procedure as before, one can derive the form of the potential, using the fact that the solutions (\ref{eq:other_sing_phi0}) satisfy the EoM. However, for the above two behaviors, it is impossible to obtain a finite action at the center of the bubble, as the integrand always blows up when $\rho \rightarrow 0$. Therefore, we exclude these two possibilities.\footnote{As mentioned before, one can introduce the cutoff of $\rho$-integral to regularize the action, see \cite{Mukhanov:2020pau,Mukhanov:2020wim,Mukhanov:2021ggy}. However, in this paper we are not interested in such a case.}
It is actually worth commenting that in the case where $\bar{\Phi}(\rho)$ exhibits a fractional power-law divergence [see Eq.~(\ref{eq:other_sing_phi0})]. 
The corresponding potential is given by 
\begin{align}
    V(\Phi) = \frac{n^2(n-2) M^2}{2(n+1)\rho_\star^2} \bigg(\frac{\Phi}{M}\bigg)^{2(n+1)/n} \;.
\end{align}
Using the above potential in the action (\ref{eq:action_zero}) together with the power-law divergent solution, we find that the action $S_0$ is finite if $0 < n < 1$. 
However, for simplicity, we focus the exponential potential case and do not consider this possibility in the rest of this paper. 

\subsection{Cuspy solution}\label{sec:cuspy_sol}
In the previous subsection we have derived a general form of the potential, for which the instanton solution and its first derivative diverge as $\rho \rightarrow 0$, but yield a finite action. 
For completeness, in this subsection, let us consider the possibility in which $\bar{\Phi}'(\rho)$ is a constant and non-vanishing at the center of the bubble. 
We call it a cusp solution.
Notice that this behavior does not satisfy the second of the Coleman boundary condition (\ref{eq:con_coleman}).

Let us consider the solution $\bar{\Phi}(\rho)$ with the following form,
\begin{align}\label{eq:sol_cuspy}
    \bar{\Phi}(\rho) = C_1 + C_2\rho +O(\rho^2) \;,
\end{align}
with $C_1$ and $C_2$ being a non-vanishing constant, where we assume that the true vacuum is located at $\bar{\Phi}(\rho = 0) = C_1$.
This implies $\bar{\Phi}'|_{\rho = 0} = C_2 \neq 0$, which violates the second of the condition (\ref{eq:con_coleman}).  
Inserting (\ref{eq:sol_cuspy}) into (\ref{eq:EOM_phi_zeroth}), we find that the potential must have the form,
\begin{align}\label{eq:potential_log}
    V(\Phi) = \Lambda_\star^4 \log\bigg|\frac{\Phi}{M_\star} - 1 \bigg|+\cdots \;,
\end{align}
where the dots are regular terms and $M_\star$ and $\Lambda_\star$ are constants. 
Note that $V(0) = 0$ and $({\rm d}V/{\rm d}\Phi)_{\Phi = 0} = -\Lambda_\star^4/M_\star$.
In order for the ansatz (\ref{eq:sol_cuspy}) to satisfy the EoM, one requires that the constants $C_1$ and $C_2$ are given by 
\begin{align}
    C_1 = M_\star \;, \qquad C_2 = - \frac{\Lambda_\star^2}{\sqrt{3}} \;,
\end{align}
where we have focused on the range $0 \leq \Phi/M_\star \leq 1$.\footnote{In the range $\Phi/M_\star > 1$ the potential is positive definite as well as its first derivative; therefore, the tunneling process cannot be realized in this regime.} 
Apparently, the solution is finite at $\rho = 0$, but it does not satisfy Coleman's boundary conditions. 
This suggests that there may exist solutions with non-trivial deformations beyond Coleman's instantons. 
However, since we only focus on singular solutions in this paper, we leave detailed studies of the cuspy solution and its deformation for future work.

\section{Anisotropic deformations}\label{sec:deform_instan}
\subsection{Small anisotropic deformation}\label{sec:small_deform}
In this subsection we analyze a small anisotropic and classical deformation around the $O(4)$-symmetric instanton. 
Let us consider the fluctuation, $\delta \Phi(x)$,  around the solution $\bar{\Phi}(\rho)$: 
\begin{align}\label{eq:ansatz}
    \Phi(x^\mu) = \bar{\Phi}(\rho) + \epsilon\, \delta \Phi(\rho, \vec{\theta}\,) \;,
\end{align}
where $\epsilon$ is the anisotropic parameter with $|\epsilon|\ll1$.
Here we emphasize that we focus on deformations that satisfy the classical equations of motion. 
Without loss of generality, $\delta \Phi$ is a function of both $\rho$ and the angular variables $\vec{\theta} \equiv\{\theta, \chi, \phi\}$. 
The potential, $V(\Phi)$, can then be expanded up to $\mathcal{O}(\epsilon^2)$ as 
\begin{align}\label{Vexpand}
  V(\Phi) = V_0 + \epsilon \frac{\de V}{\de \Phi}\bigg|_{\bar{\Phi}} \delta \Phi + \frac{\epsilon^2}{2} \frac{\de^2V}{\de\Phi^2}\bigg|_{\bar{\Phi}} \delta \Phi^2 + \cdots \;.  
\end{align}
We will see, in the next subsection, that in order to obtain a finite and non-trivial contribution to the on-shell action up to $\mathcal{O}(\epsilon^2)$, the second derivative of the potential has to satisfy a certain condition. Inserting \eqref{eq:ansatz} and \eqref{Vexpand} into \eqref{eq:EOM_phi}, the EoM for $\delta \Phi(\rho, \vec{\theta}\,)$ is expressed as
\begin{align}\label{eq:Phi_1}
        \frac{1}{\rho^{3}} \partial_\rho \big(\rho^{3}\partial_\rho\delta \Phi\big) + \frac{1}{\rho^2} \Delta_{S^{3}} \delta \Phi - \frac{\de^2 V}{\de\Phi^2}\bigg|_{\bar{\Phi}} \delta \Phi = 0\;,
\end{align}
with $\Delta_{S^{3}}$ being the Laplace operator on the 3-sphere:
\begin{align}
     \Delta_{S^{3}} \equiv \frac{1}{\sin^2\theta}\,\partial_\theta\big(\sin^2 \theta\,\partial_\theta \big) + \frac{1}{\sin^2\theta} \bigg[\frac{1}{\sin\chi}\,\partial_\chi \big(\sin\chi\,\partial_\chi \big) + \frac{1}{\sin^2\chi}\,\partial_\phi^2  \bigg] \;.
\end{align}
Note that $\de^2 V/\de \Phi^2$ evaluated on $\bar{\Phi}$ is only a function of $\rho$. 
The fact that Eq.~(\ref{eq:Phi_1}) is linear in $\delta \Phi$ allows us to solve for the solution using the separation of variables.  
Namely, the solution $\delta \Phi(\rho, \vec{\theta}\,)$ can be decomposed as
\begin{align}
    \delta \Phi(\rho, \vec{\theta}\,) = \sum_{L,M} A_{LM}(\rho)\,Y_L^M(\vec{\theta}\,) \;, \label{eq:sol_PHi1}
\end{align}
where the function $Y_L^M(\vec{\theta}\,)$ is the 3-dimensional spherical harmonics \cite{doi:10.1063/1.527513} which satisfies 
\begin{align}
    \Delta_{S^{3}} Y_L^M(\vec{\theta}\,) = -L(L + 2) Y^M_L(\vec{\theta}\,) \;,
\end{align}
with $M$ being a multi-index characterizing the magnetic quantum numbers $\{m_\phi, m_\chi, m_\theta\}$ satisfying $|m_\phi| \leq m_\chi \leq L \equiv m_\theta$.
We refer the reader to appendix~\ref{app:spherical_harmonics} for general properties of the $(n-1)$-dimensional spherical harmonics.

For simplicity, we consider the case where fluctuation $\delta \Phi$ is only a function of $\rho$ and $\theta$, i.e., $m_\phi = m_\chi = 0$. Thus, we have
\begin{align}
    Y_L(\theta) &= 
    \frac{{}_{2}\bar{P}_{m_\chi = 0}^{m_\phi = 0}(\chi)\,{}_{3}\bar{P}_{m_\theta = L}^{m_\chi = 0}(\theta)}{\sqrt{2 \pi}} 
  = \frac{\sin[(L+1)\theta]}{\sqrt{2}\,\pi\sin\theta} \;, \label{eq:Y_theta}
    \end{align}
where we have used the formulas (\ref{eq:n_spherical_H})--(\ref{eq:asso_Legen}) and $P_0^0(\cos\theta) = 1$. 
We note that in the limits $\theta \rightarrow 0$ and $\pi$, we have $Y_L \rightarrow (L+1)/(\sqrt{2}\,\pi)$
and $(-1)^L(L+1)/(\sqrt{2}\,\pi)$, respectively.

The radial component $A_L(\rho)$ in Eq.~(\ref{eq:sol_PHi1}) satisfies
\begin{align}\label{eq:A(rho)}
    A''_L + \frac{3}{\rho} A'_L - \bigg[ \frac{\de ^2 V}{\de\Phi^2}\bigg|_{\bar{\Phi}} + \frac{L(L + 2)}{\rho^2} \bigg]A_L = 0 \;,
\end{align}
where we have omitted the index $M$. 
We will see that this equation, in fact, will lead to a condition on the potential in order for the solution $\delta \Phi$ to give a finite and non-trivial contribution to the on-shell action.
Before discussing general on-shell actions, let us first consider the cases of monopole ($L = 0$) and dipole ($L=1$) modes.

First of all, in the case of $L=1$, we can identify one of the independent solution with a coordinate gauge transformation. To see this, let us consider an infinitesimal coordinate transformation $x^\mu \rightarrow x^\mu + \varepsilon^\mu$. First consider the Cartesian coordinates, $\mu=(\tau,x,y,z)$. Then it is apparent that $\varepsilon^\mu$ are constants. 
Transforming them to those in the spherical coordinates $(\rho,\theta,\chi,\phi)$, we find
\begin{equation}
\begin{aligned}
    \varepsilon^\rho &= \varepsilon^z \cos\theta + \sin\theta [\,\varepsilon^y \cos\chi + (\varepsilon^x \cos\phi + \varepsilon^\tau \sin\phi)\sin\chi] \;, \\
    \varepsilon^\theta &= -\frac{1}{\rho} \big\{\varepsilon^z \sin\theta - \cos\theta [\,\varepsilon^y \cos\chi + (\varepsilon^x \cos\phi + \varepsilon^\tau \sin\phi)\sin\chi] \big\} \;, \\
    \varepsilon^\chi &= -\frac{\sin\chi}{\rho\sin\theta} \big[\,\varepsilon^y - \varepsilon^x \cos\phi \cot\chi - \varepsilon^\tau \cot\chi\sin\phi \big] \;, \\
    \varepsilon^\phi &= \frac{\varepsilon^\tau \cos\phi - \varepsilon^x \sin\phi}{\rho \sin\theta\sin\chi}  \;, 
\end{aligned}
\end{equation}
where we have used the transformations,
\begin{equation}
\begin{aligned}
    \tau = \rho \sin\theta \sin\chi \sin\phi \;,  \qquad x = \rho \sin\theta \sin\chi \cos\phi \;, \qquad y =  \sin\theta \cos\chi \;, \qquad z = \rho \cos\theta \;. 
\end{aligned}
\end{equation}

Let us focus on the case $\varepsilon^z\neq0$ while $\varepsilon^\tau=\varepsilon^x=\varepsilon^y=0$ for simplicity.\footnote{ 
The extension to the general case is straightforward, but since it only introduces an inessential complication, we do not discuss it here.}
Expressing the parameters $\varepsilon^\mu$ in the spherical coordinates as
\begin{equation}\label{eq:epsilon_rho}
    \begin{aligned}
    \varepsilon^\rho = \sum_{L,M} T_{LM}(\rho) Y^M_L(\vec{\theta}) \;, \qquad
    \varepsilon^a = \sum_{L,M} \Theta_{LM} (\rho) \bar{\nabla}^a Y^M_L(\vec{\theta}) \;, 
\end{aligned}
\end{equation}
where the index $a = \{\theta, \chi, \phi\}$ and $\bar{\nabla}^a$ denotes the covariant derivative with respect to the 3-sphere metric, we immediately find that
\begin{align}
    T_0 = 0 \;, \qquad T_1 = \frac{\pi \varepsilon^z}{\sqrt{2}} \;, \qquad T_{L \geq 2} = 0 \;.
\end{align}
Therefore, the variable $A_L(\rho)$ transforms as 
\begin{equation}\label{eq:gauge_transf}
\begin{aligned}
    A_0 \rightarrow A_0 \;, \qquad A_1 \rightarrow A_1 - \frac{\pi \varepsilon^z}{\sqrt{2}}  \bar{\Phi}' \; \qquad 
     A_{L \geq 2} \rightarrow A_{L \geq 2} \;.
\end{aligned}
\end{equation} 
It is easy to see that one of the solutions for $A_1$ is proportional to $\bar{\Phi}'$. 
Hence, one can choose $\varepsilon^z$ such that the dipole mode $A_1$ vanishes. This means it is a gauge mode.
Notice that this is the case only for regular deformations. 
As we will discuss later, a singular dipole deformation exists, but it gives infinite action. 
Thus, we disregard the regular dipole deformation.
Notice that from Eq.~(\ref{eq:gauge_transf}) the modes with $L = 0$ and $L \geq 2$ transform to itself under the gauge transformation. That is, they are gauge-invariant.

In the case of $L=0$, i.e., for monopole deformations, there should not be physically meaningful solutions.
The reason is that since the background $\bar{\Phi}$ is $O(4)$-symmetric, there does not exist an infinitesimal $O(4)$-symmetric deformation of $\bar{\Phi}$ that can also be a solution, except for systems that have the conformal symmetry which is not the case of our interest. 
In other words, $\bar{\Phi}(\rho) + \epsilon A_{L = 0}(\rho)$ cannot satisfy boundary conditions that are imposed to find $O(4)$-symmetric instanton solutions.
In fact, we will see in the next section that the finite action requirement forbids the existence of such a solution.
Therefore, we exclude the monopole deformations in our analysis. 
For the rest of the paper, we only focus on the modes with $L\geq2$.

In the next subsection, we will analyze the contributions coming from $\delta \Phi$ to the total on-shell action. 
In particular, we will obtain a generic behavior of $\delta \Phi$ in order for the action to be finite.    

\subsection{On-shell action up to $\mathcal{O}(\epsilon^2)$}\label{sec:bounda_singular}
Let us expand the action (\ref{eq:action_Eu}) using the ansatz (\ref{eq:ansatz}) up to second order in $\epsilon$. We then have 
\begin{widetext}
\begin{equation}
S =  \int \mathrm{d}^4x \sqrt{g}~\bigg\{\frac{1}{2}\bar{\Phi}'^2 + V(\bar{\Phi}) + \epsilon \bigg[\partial_\mu \bar{\Phi} \partial^\mu \delta \Phi + \frac{\de V}{\de\Phi}\bigg|_{\bar{\Phi}} 
\delta \Phi \bigg]  + \epsilon^2\bigg[\frac{1}{2}(\partial_\mu \delta \Phi)^2 + \frac{1}{2}\frac{\de^2V}{\de\Phi^2}\bigg|_{\bar{\Phi}} \delta \Phi^2 \bigg] \bigg\} \;.
\end{equation}
\end{widetext}
It is then straightforward to show that the action at first order in $\epsilon$ vanishes due to the zeroth-order equation of motion [Eq.~(\ref{eq:EOM_phi_zeroth})] after performing an integration by parts.
We note that the $\mathcal{O}(\epsilon)$ boundary term vanishes since it behaves as $\rho^3 \bar{\Phi}' \delta\Phi$ in the $\rho\to0$ limit under the condition that $\delta\Phi$ is regular at $\rho = 0$.
Therefore, we are left with 
\begin{align}\label{eq:action_Phi1}
S = \int \mathrm{d}^4x \sqrt{g}\,\bigg\{\frac{1}{2}\bar{\Phi}'^2 + V(\bar{\Phi}) 
+ \frac{\epsilon^2}{2}\bigg[(\partial \delta \Phi)^2 + \frac{\de^2V}{\de\Phi^2}\bigg|_{\bar{\Phi}} \delta \Phi^2 \bigg] \bigg\} \equiv S_0 + \epsilon^2 S_2 \;,     
\end{align}
where the action $S_0$ refers to the one associated with $\bar{\Phi}$ and $S_2$ denotes the one coming from the anisotropic fluctuation $\delta \Phi$. 
Note that the effective mass for $\delta \Phi$ that is given by $\de ^2V/\de\Phi^2$ evaluated on $\bar{\Phi}(\rho)$ is only a function of $\rho$; the only non-trivial angular dependence (apart from the one in the volume element $\sqrt{g}$) appears through the fluctuation $\delta \Phi(\rho, \vec{\theta}\,)$. 
Then, we insert the solution (\ref{eq:sol_PHi1}) into the action $S_2$ to obtain
\begin{align}
    S_2 = \sum_{L \geq 2} \int \mathrm{d}\rho\, \mathcal{L}_{2L} \label{eq:second_actio_AL} \;,
\end{align}
where
\begin{align}\label{eq:integrand_A_L}
    \mathcal{L}_{2L} \equiv \frac{\rho^3}{2} \bigg\{A_L'^2 + \bigg[\frac{L(L+2)}{\rho^2}
    + \frac{\de^2 V}{\de \Phi^2}\bigg|_{\bar{\Phi}}\bigg] A_L^2\bigg\} \;,
\end{align}
and we have used the orthogonality condition (\ref{eq:orth_Y}) for spherical harmonics.

In fact, for a regular deformation, Eq.~(\ref{eq:second_actio_AL}) can be rewritten as
\begin{align}
   S_2 = \pi^2 \sum_{L \geq 2} \rho^3 A_L(\rho) A'_L(\rho) \bigg|_{\rho = 0}^{\rho = \infty} \;, \label{eq:s_2_phi_1}
\end{align}
where we have performed an integration by parts and used Eq.~(\ref{eq:A(rho)}).
We see that the action $S_2$ only receives contributions coming from the behavior of $A_L$ at the origin and infinity.
Since in this case $A_L$ is regular everywhere, the action $S_2$ is finite.
We emphasize that Eq.~(\ref{eq:s_2_phi_1}) holds true only for regular deformations. 

Near the origin where the potential is exponential, we find the exact solution for $A_L$ from Eq.~(\ref{eq:A(rho)}),
\begin{equation}
    A_L(\rho) =a_1\rho^{-1+\sqrt{j}}+a_2\rho^{-1-\sqrt{j}}\,.
\end{equation}
where $a_1$, $a_2$ are constants and $j \equiv (L - 1)(L + 3)$ ($\geq5$ for $L\geq2$). 
We see that the above solution either diverges as $\rho \to 0$ when $a_2 \neq 0$ or it is regular at ${\rho} = 0$ when $a_2 = 0$. 

Let us first consider the solution with the singular behavior. 
Note the singular $L=1$ solution cannot be removed by a gauge transformation.
For $L = 1$ the two solutions of Eq.~(\ref{eq:A(rho)}) degenerate. In this case the singular solution is given by $A_{1,\rm sgr}\propto{\rho}\log({\rho})$. 
This solution leads to logarithmic divergence of the action. 
Therefore, we disregard the singular dipole deformations.
For $L \geq 2$, the singular solution $A_{L,\rm sgr}\propto {\rho}^{-1 -\sqrt{j}}$ lead to power-law divergence, unless one introduces the UV cutoff. 
Note that, as discussed before in section~\ref{sec:bounda_singular}, in order for the second-order action (\ref{eq:second_actio_AL}) evaluated on the classical solution $\delta \Phi$ to be finite and non-vanishing the singular mode function $A_L(\rho)$ must behave as $\rho^{-1/2}$, which cannot be realized for $L \geq 1$.
Therefore, a non-trivial singular deformation around the $O(4)$-symmetric singular instanton solutions is not allowed. 

Let us turn to the solution with the regular behavior $A_{L,\rm reg}\propto {\rho}^{-1+\sqrt{j}}$. 
In this case, the action is finite provided that the solution decays exponentially as ${\rho} \to \infty$. 
This means that one is required to solve for $A_L$ in the $\Phi<0$ region with the boundary condition that it decays exponentially at ${\rho}\to\infty$ and match it to the solution $A_L = a_1 {\rho}^{-1 + \sqrt{j}}$ in the exponential regime.
Apparently this cannot be always done. In other words, the existence of a regular solution depends on the parameters of the potential. 
Namely, we have to scan all the possible values of the model parameters that allow the existence of singular $O(4)$-symmetric instantons with finite action and see if they allow regular deformations.

We note that if $A_L \propto \rho^{1/2}$, the second-order action (\ref{eq:s_2_phi_1}) gives a non-vanishing value. 
However, as we have seen above all regular solutions have a power greater than $1/2$ for $L \geq 2$. 
Hence, the second-order action vanishes for any regular deformations. 

\section{Concrete examples}\label{sec:con_example}
In this section, we examine two specific examples: The false vacuum side of the potential given by a cubic function (section~\ref{sec:cubic_potential}) and thate by a piecewise quadratic function (section~\ref{sec:piecewise_qud}).
For both cases, we find the existence of singular $O(4)$-symmetric instanton solutions. However, we find that, while the cubic potential does not allow for any small deformations with finite action, the piecewise quadratic potential does.

\subsection{Cubic potential}\label{sec:cubic_potential}
Let us now consider the following potential:
\begin{align}\label{eq:potential}
    \frac{V(\Phi)}{\alpha^4} = \left\{
\begin{matrix}
-\Phi_\star^4 \exp(2 \Phi/\Phi_\star)\;, & \Phi \geq 0 \\
-a \Phi^3 - b \Phi^2 - c\Phi - d \;, & \Phi \leq 0 
\end{matrix}
\right. \;,
\end{align}
where $a$, $b$, $c$ and $d$ are constants which will be fixed later, and $\alpha$ is an overall dimensionless parameter.
Notice that we put a minus sign in front of all the terms since they will change to positive sign when entering into the equation of motion (\ref{eq:EOM_phi_zeroth}).
From the above potential, we choose $\Phi = 0$ to be the location at which the potential changes its behavior. 
Here we use $\Phi_\star$ to denote a pivot scale for a scalar field.
Note that in this example we have only assumed the form of the cubic potential, whereas the exponential one is a form consistent with the singular instanton whose action is finite.

Before solving the zeroth-order equation for $\bar{\Phi}$, let us analyze the behavior of the potential (\ref{eq:potential}).
Using continuity conditions of the potential and its first derivative at $\Phi = 0$, we obtain 
\begin{align}
    c = 2 \Phi_\star^3 \;, \quad d = \Phi_\star^4 \;.
\end{align}
Substituting $c$ and $d$ in the cubic potential we have 
\begin{align}\label{eq:V_3}
    \frac{V(\Phi \leq 0)}{\alpha^4} = -a \Phi^3 - b \Phi^2 - 2 \Phi_\star^3 \Phi - \Phi_\star^4 \;.
\end{align}
We see that after imposing the continuities of the potential and its derivative at $\Phi = 0$ we are left with only two free parameters, namely $a$ and $b$. 
Also, the two extrema of the potential (\ref{eq:V_3}) are located at
\begin{align}\label{eq:local_max_min_Phi}
    \Phi_{\pm}(a,b, \Phi_\star) = -\frac{b}{3a} \pm \frac{1}{3a}\sqrt{b^2 - 6 a \Phi_\star^3 } \;.
\end{align}
Clearly, the two extrema coincide when $b^2 = 6 a \Phi_\star^3 $, i.e., they are neither local maximum nor local minimum. 
The existence of the real and negative $\Phi_{\pm}$ requires that 
\begin{align}\label{eq:con_real_negative_Phi}
    b^2 > 6 a \Phi_\star^3 \;, \quad \Phi_{\pm} < 0 \;.
\end{align}
Notice that the second requirement comes from the fact that the cubic potential is only present in the regime $\Phi \leq 0$.
Moreover, demanding that $\Phi_+$ and $\Phi_-$ are the local minimum and local maximum respectively yields the conditions:
\begin{align}\label{eq:con_second_V}
    -\frac{\de^2 V}{\de \Phi^2}\bigg|_{\Phi = \Phi_+} > 0 \;, \quad -\frac{\de^2 V}{\de \Phi^2}\bigg|_{\Phi = \Phi_-} < 0 \;.
\end{align}
In fact, the conditions (\ref{eq:con_real_negative_Phi}) and (\ref{eq:con_second_V}) give rise to the allowed region of parameters $a$ and $b$, as shown by the shaded green region in Fig.~\ref{fig:Parameter_space_a_b}.
As we will see below, for each value of $a$ (or $b$) there exists a corresponding value of $b$ (or $a$) such that the singular instanton solution exists. These are shown by the red data points in Fig.~\ref{fig:Parameter_space_a_b}.
\begin{figure}[t]
    \centering
    \includegraphics[width=0.5\linewidth]{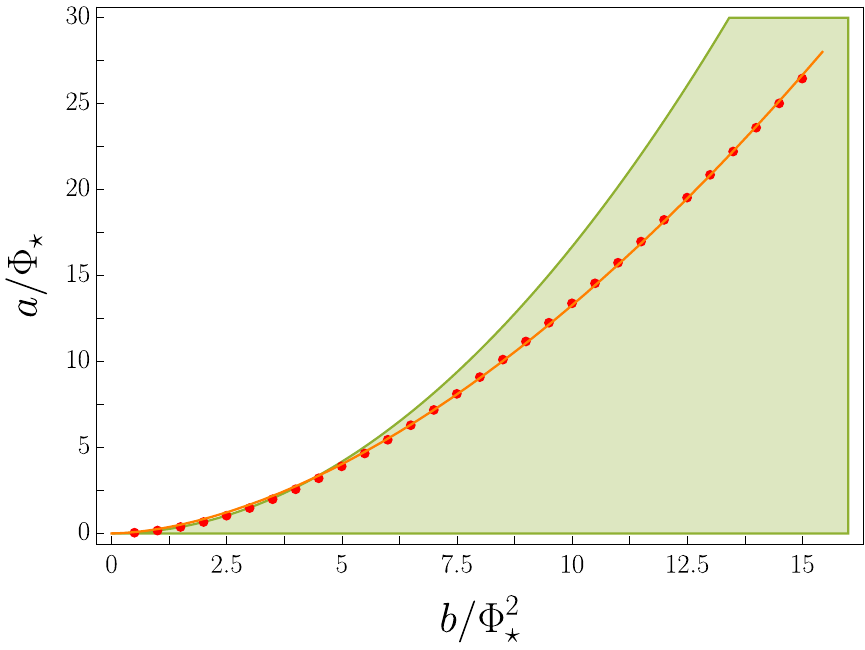}
    \caption{~Parameter space of $a$ and $b$ of the potential (\ref{eq:V_3}). The shaded green region illustrates the allowed region consistent with the conditions (\ref{eq:con_real_negative_Phi}) and (\ref{eq:con_second_V}).
    The red data points and the orange line represent respectively the numerical values of $a$ and $b$ and their fitting relation, for which there exist the singular instanton solutions.
    The fitting formula is approximately given by $a/\Phi_\star = 0.25\,(b/\Phi_\star^2)^{1.72}$.}
    \label{fig:Parameter_space_a_b}
\end{figure}
Additionally, we plot the potential (\ref{eq:V_3}) for several values of $a$ and $b$ in Fig.~\ref{fig:cubic_potential}. 
The red solid line refers to the exponential potential when $\Phi \geq 0$, while for $\Phi \leq 0$ the blue dot-dashed, the green dashed and the pink dotted curves correspond to the choices $\{a = 9.09\Phi_\star, b = 8\Phi_\star^2\}$, $\{a = 13.38\Phi_\star, b = 10\Phi_\star^2\}$ and $\{a = 18.24\Phi_\star, b = 12\Phi_\star^2\}$ respectively.
Note that the values of $a$ and $b$ in Fig.~\ref{fig:cubic_potential} are those which ensure the existence of singular instanton solutions.
The black point represents the matching location at $\Phi = 0$.
We see in Fig.~\ref{fig:cubic_potential} that as we increase $a$ (or $b$) along the orange line of Fig.~\ref{fig:Parameter_space_a_b} the value of the false vacuum $\Phi_-$ becomes smaller, while $-V(\Phi_-)$ becomes larger.
This in fact explains the behavior of the bounce action, as we will compute below.

\begin{figure}[t]
\includegraphics[width=0.5\textwidth]{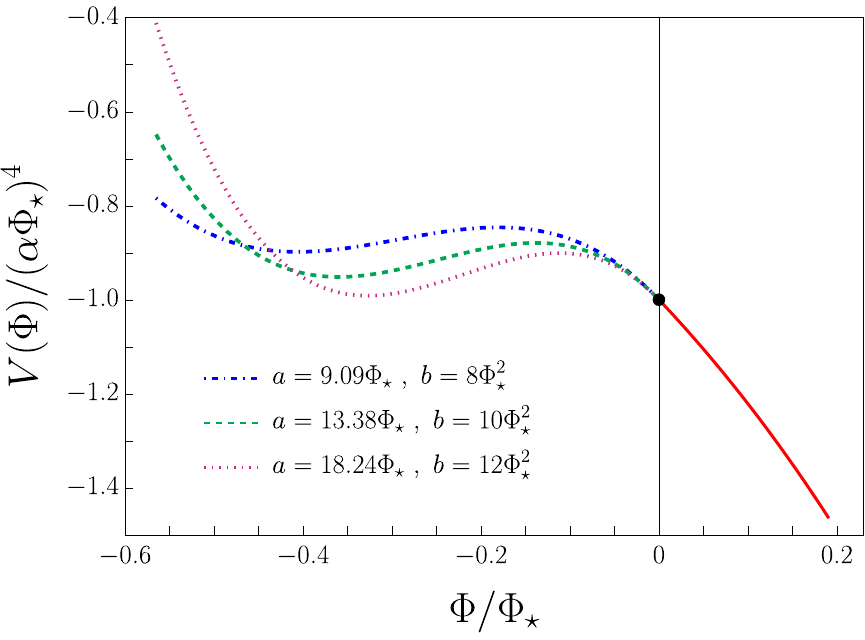}
\caption{~The potential (\ref{eq:potential}) as a function of $\Phi/\Phi_\star$. 
For $\Phi \leq 0$, the blue dot-dashed, the green dashed and the pink dotted lines correspond to the potential (\ref{eq:V_3}) with $\{a = 9.09\Phi_\star, b = 8\Phi_\star^2\}$, $\{a = 13.38\Phi_\star, b = 10\Phi_\star^2\}$ and $\{a = 18.24\Phi_\star, b = 12\Phi_\star^2\}$ respectively, for which their corresponding false vacua are located at $\Phi_- = -0.40\Phi_\star$, $\Phi_- = -0.36\Phi_\star$ and $\Phi_- = -0.33\Phi_\star$.
These values of $a$ and $b$ ensure the existence of the singular instantons, shown in Fig.~\ref{fig:Parameter_space_a_b}.
The red solid line represents the exponential potential in the range $\Phi \geq 0$.
Additionally, the black dot denotes the matching location at $\Phi = 0$.}
\label{fig:cubic_potential} 
\end{figure}

Let us first consider the $O(4)$-symmetric solutions $\bar{\Phi}$. In the exponential regime ($\Phi \geq 0$) we have 
\begin{align}\label{eq:sol_expo}
    \bar{\Phi}(\tilde{\rho}) = -\Phi_\star\log(\tilde{\rho}) \;, \qquad  0 \leq \tilde{\rho} \leq 1
\end{align}
with $\tilde{\rho} \equiv \alpha^2 \Phi_\star \rho$. 
Notice that here we only focus on the singular solution as $\rho \rightarrow 0$, so that we disregard regular behavior.\footnote{It should be noted that once the boundary conditions, $\bar{\Phi} = \Phi_{\rm FV}$ as $\rho \rightarrow \infty$ and $\Phi'_0 \rightarrow \infty$ at $\rho = 0$, are imposed, the solution for $\bar{\Phi}$ is unique and is given by Eq.~(\ref{eq:sol_expo}) connected with that obtained in the cubic-potential regime.} 
It is important to note that, although this solution is singular as $\rho \rightarrow 0$, it gives rise to a finite action, as discussed in section~\ref{sec:gen_potential}.\footnote{Note that in our case there is no need to introduce the UV cutoff, unlike in \cite{Mukhanov:2021ggy} where a UV cutoff was introduced to make the instanton action finite, determining the region of the so-called quantum core.}
Actually, in appendix~\ref{app:sextic_solution} we show that our model can be regarded as the singular limit of a series of regular potentials with regular instanton solutions with finite action.

\begin{figure}[t]
    \centering
    \includegraphics[width=0.5\linewidth]{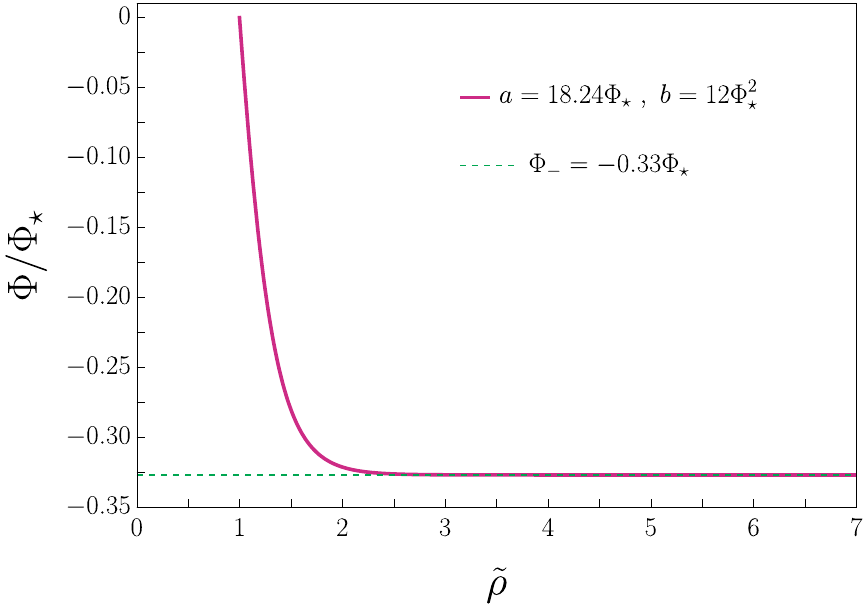}
    \caption{~The numerical instanton solution (pink solid line) in the cubic regime ($\Phi < 0$) with $a = 18.24\Phi_\star$ and $b = 12\Phi_\star^2$. The green dashed line represents the value of $\Phi$ at the false vacuum. Here we use $\alpha = 1$.}
    \label{fig:cubic_num_solution}
\end{figure}

In the cubic-potential regime ($\Phi \leq 0$) we numerically solve Eq.~(\ref{eq:EOM_phi_zeroth}), subjected to the boundary conditions:
\begin{align}
    \bar{\Phi} |_{\tilde{\rho} = \tilde{\rho}_{\rm f}} = \Phi_-(a,b) \;, \qquad \frac{\mathrm{d}\bar{\Phi}}{\mathrm{d}\tilde{\rho}} \bigg|_{\tilde{\rho} = 1} = - \Phi_\star \;,
\end{align}
where $\tilde{\rho}_{\rm f}$ denotes the final point of integration. 
The second condition, which comes from the solution (\ref{eq:sol_expo}), is imposed to smoothly connect with the exact solution (\ref{eq:sol_expo}) in the regime $\Phi \geq 0$.
We obtain the numerical instanton solutions in the regime $\tilde{\rho} \geq 1$ with several values of $a$ and $b$. 
In particular, we find that the solutions exist for specific pairs of parameters $a$ and $b$, shown by the red points in Fig.~\ref{fig:Parameter_space_a_b}.
It should be noted that for small values of $a$ and $b$ the value of $\tilde{\rho}_{\rm f}$ must be sufficiently large such that the field approaches the false vacuum $\Phi_-$. 
In our numerical computation we use $\tilde{\rho}_{\rm f} = 20$ for $b < 3$, while we use $\tilde{\rho}_{\rm f} = 7$ for $b \geq 3$.
Also, we give a simple fitting formula (orange line in Fig.~\ref{fig:Parameter_space_a_b}) between parameters $a$ and $b$: 
$a/\Phi_\star = 0.25\,(b/\Phi_\star^2)^{1.72}$.
If, on the other hand, the parameters $a$ and $b$ are not related by the above relation, one then obtains either undershoot or overshoot solution.
In Fig.~\ref{fig:cubic_num_solution} we show the numerical solution with $a = 18.24\Phi_\star$ and $b = 12\Phi_\star^2$. 
We see in Fig.~\ref{fig:cubic_num_solution} that for sufficiently large $\tilde{\rho}$ the instanton approaches the false vacuum $\Phi_-$ (green dashed line), whereas as $\tilde{\rho} \rightarrow 1$ the solution diverges logarithmically as expected. 
It is important to note that we have not used the thin-wall approximation to solve for the instanton solution $\bar{\Phi}$.
Moreover, it is worth pointing out that for those parameters slightly off the allowed region, we would obtain either a regular instanton or no instanton solution.

Let us evaluate the zeroth-order action. Written in terms of all the dimensionless variables, the zeroth-order bounce action reads
\begin{align}\label{eq:action_eva_example}
    B_0[\bar{\Phi}] = \frac{2 \pi^2}{\alpha^4} \int_0^{\tilde{\rho}_{\rm f}}\mathrm{d}\tilde{\rho}~\tilde{\rho}^3  \bigg[\frac{1}{2\Phi_\star^2}\bigg(\frac{\mathrm{d}\bar{\Phi}}{\mathrm{d}\tilde{\rho}}\bigg)^2 + \frac{V(\bar{\Phi}) - V(\Phi_{\rm FV})}{(\alpha \Phi_\star)^4}\bigg] \;,
\end{align}
where the factor $2\pi^2$ is obtained from the integral over the angular variables and $V(\Phi_{\rm FV})$ denotes the potential at the false vacuum, i.e., $\Phi_{\rm FV} = \Phi_-$.
Note that in (\ref{eq:action_eva_example}) we have subtracted the contribution coming from the false vacuum, so that $B_0$ vanishes when $\bar{\Phi} = \Phi_-$ and $\mathrm{d}\bar{\Phi}/\mathrm{d}\tilde{\rho} = 0$ at $\tilde{\rho} = \tilde{\rho}_{\rm f}$.

\begin{figure}[t]
    \centering
    \includegraphics[width=0.5\linewidth]{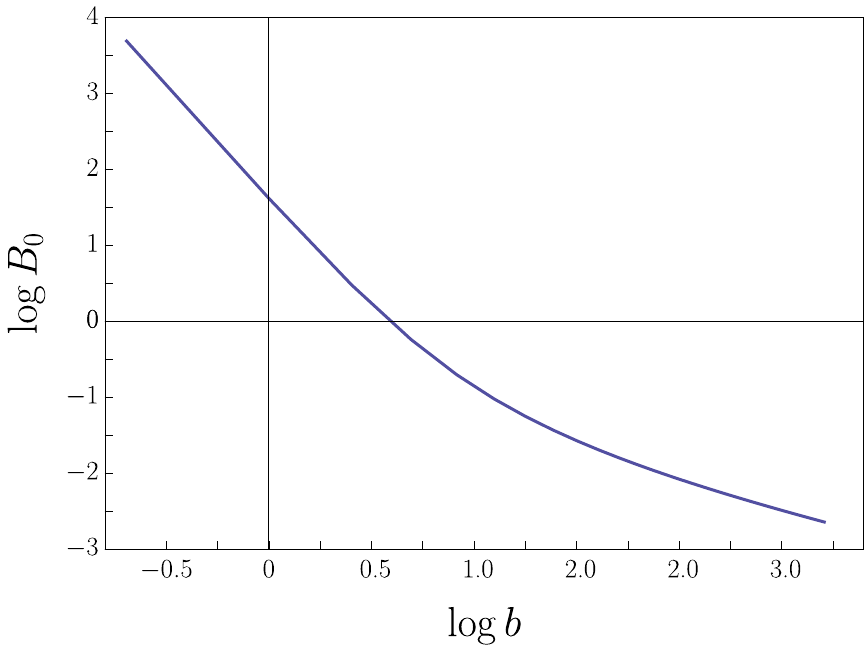}
    \caption{~The bounce action (\ref{eq:action_eva_example}) evaluated on the singular instantons as a function of $b$, shown by the blue solid line.}
    \label{fig:total_action_example}
\end{figure}

We evaluate the action (\ref{eq:action_eva_example}) numerically and the result is shown in Fig.~\ref{fig:total_action_example}.
Note that we perform the numerical integration over the range $\tilde{\rho} \in [0,\tilde{\rho}_{\rm f}]$ with $\tilde{\rho}_{\rm f} = 20$ for $b < 3$ and $\tilde{\rho}_{\rm f} = 7$ for $b \geq 3$.
In Fig.~\ref{fig:total_action_example} we show the numerical values of the action (\ref{eq:action_eva_example}), evaluated on the instanton solutions, with respect to the parameter $b$. 
Notice that the existence of the singular instantons is characterized by the orange line in Fig.~\ref{fig:Parameter_space_a_b}.

Let us comment on the behavior of the bounce action $B_0$.
First, we see that in the plot the bounce action $B_0$ decreases as increasing $b$.
The reason is as follows. 
When both $a$ and $b$ increase (along the orange line of Fig.~\ref{fig:Parameter_space_a_b}) the value of $-V(\Phi_-)$ is becoming larger, see Fig.~\ref{fig:cubic_potential}.
Since the potential (\ref{eq:potential}) enters into the bounce action with the negative sign, we therefore obtain a lower value of $B_0$ as increasing both $a$ and $b$. 
This implies that the decay rate becomes larger, corresponding to the process with higher proability.
Second, one can analytically obtain the bounce action in the exponential regime and show that it gives negative contribution.
This is not surprising due to the unbounded exponential potential. 
However, the total bounce action remains positive since the exponential part is subdominant compared to the cubic-potential part, which is positive definite. 
In other words, most of the tunneling process happens in the regime of cubic potential. 
Therefore, the zeroth-order solutions we found in this example are stable (positive bounce action).

For the rest of this section, we discuss the existence of the solution for deformations around $\bar{\Phi}$. 
For later convenience, we introduce a new variable $f_L$ via $f_L \equiv \tilde{\rho}^{3/2}A_L/\Phi_\star$. 
Therefore, Eq.~(\ref{eq:A(rho)}) becomes 
\begin{align}\label{eq:EoM_fluctuation}
    \frac{\de^2 f_L}{\de \tilde{\rho}^2} - \bigg[\frac{1}{\alpha^4 \Phi_\star^2} \frac{\de^2 V}{\de \Phi^2}\bigg|_{\bar{\Phi}} + \frac{3 + 4L (L + 2)}{4\tilde{\rho}^2} \bigg]f_L = 0 \;.
\end{align}
For the solution to be regular, we require the boundary conditions,
\begin{align}\label{eq:sol_f_expo}
    f_L(\tilde{\rho}) \propto \tilde{\rho}^{\frac{1}{2} + \sqrt{j}} \;; \qquad \tilde{\rho} \to 0 \;,
\end{align}
and $f_L(\tilde{\rho})$ vanishes exponentially as $\tilde{\rho} \to \infty$.

In this example, from the potential (\ref{eq:potential}) we have 
\begin{align}
   \frac{1}{\alpha^4}\frac{\de^2 V}{\de \Phi^2}\bigg|_{\bar{\Phi}} = 
   \left\{
\begin{matrix}
-4 \Phi_\star^2 \exp(2 \bar{\Phi}/\Phi_\star)\;, & \Phi \geq 0 \\
-6 a \bar{\Phi} - 2 b  \;, & \Phi \leq 0 
\end{matrix}
\right. 
   \;.
\end{align}
We numerically scanned all of the values of $a$ and $b$ (along the orange curve of Fig.~\ref{fig:Parameter_space_a_b}), for which the background solution exists, but none of them could match the exponentially decaying mode at infinity. 
Therefore, in this case, we conclude that there exits no non-trivial regular deformed instanton solution with finite action. 

\subsection{Piecewise quadratic potential}\label{sec:piecewise_qud}
In this subsection, we study another example of the potential. It is given by a piece-wise quadratic potential, 
\begin{align}\label{eq:piecewise_potential}
    \frac{V(\Phi)}{\alpha^4} = \left\{
\begin{matrix}
-\Phi_{\star}^4 \exp(2 \Phi/\Phi_\star)\;, & \Phi \geq 0 \\
V_{1}(\Phi) \;, & \Phi_2 \leq \Phi \leq 0 \\
V_{2}(\Phi) \;, & \Phi \leq \Phi_2 
\end{matrix}
\right. \;,
\end{align}
where 
\begin{align}
    V_{1}(\Phi) &\equiv -\frac{1}{2}m_1^2(\Phi - \Phi_{\rm P})^2 - \Lambda_1^4 \;, \\
    V_{2}(\Phi) &\equiv \frac{1}{2}m_2^2(\Phi - \Phi_{\rm M})^2 - \Lambda_2^4 \;, 
\end{align}
where $\Lambda$, $m_1$, $m_2$, $\Phi_{\rm P}$, $\Phi_{\rm M}$, $\Lambda_1$, $\Lambda_2$, $\Phi_\star$ are parameters with mass dimension,
$\alpha$ is an overall dimensionless parameter.
We mention that the main result of this subsection was already presented in \cite{Sasaki:2024vul}. Here we provide additional details of the analysis.

\begin{figure}[t]
\includegraphics[width=0.5\textwidth]{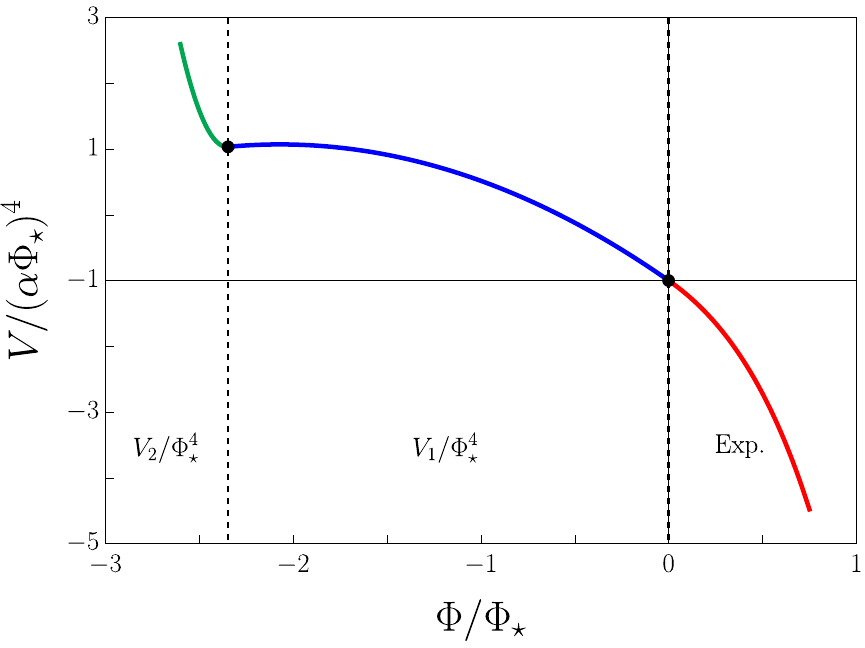}
\caption{~The potential $V(\Phi)/(\alpha\Phi_\star)^4$ [Eq.~(\ref{eq:piecewise_potential})] as a function of $\Phi/\Phi_\star$ with $m_1 = 0.98\Phi_\star$, $m_2 = 7.11 \Phi_\star$ and $\Phi_{\rm M} = -2.35\Phi_\star$. 
We choose these values of parameters such that the matching conditions for Eqs.~(\ref{eq:bg_exp})--(\ref{eq:sol_phi_0_V2}) are satisfied.
The red line represents the exponential potential, while the blue and the green lines refer to the potentials $V_1$ and $V_2$ in Eq.~(\ref{eq:piecewise_potential}) respectively. The vertical dashed lines in the plot represent the two matching locations at $\Phi = 0$ and at $\Phi = \Phi_2 = -2.34\Phi_\star$, from right to left. The false vacuum for this particular choice of parameters is located at $\Phi_{\rm M} = -2.35\Phi_\star$.}
\label{fig:piecewise_potential} 
\end{figure}

We use $\Phi_{\rm P}$ and $\Phi_{\rm M}$ to denote the locations of the maximum and the minimum (false vacuum) in $V_1$ and $V_2$, respectively.
We then impose the continuity conditions for the potential and its first derivative at $\Phi = 0$ and $\Phi = \Phi_2$. 
By doing so, the parameters $\{\Lambda_1, \Lambda_2, \Phi_{\rm P}, \Phi_2\}$ can be fixed in terms of $\{m_1, m_2 ,\Phi_{\rm M}\}$ as
\begin{equation}\label{paracon}
\begin{aligned}
     \Phi_{\rm P} &= -\frac{2\Phi_\star^3}{m_1^2} \;, \qquad \Phi_2 = -\frac{2 \Phi_\star^3}{m_1^2 + m_2^2}\bigg(1 - \frac{m_2^2\Phi_{\rm M}}{2\Phi_\star^3}\bigg) \;, \\
    \Lambda_1^4 &=  \Phi_\star^4 \bigg(1 - \frac{2\Phi_\star^2}{m_1^2}\bigg) \;, \qquad 
    \Lambda_2^4 = -\frac{2\Phi_\star^6 }{m_1^2+m_2^2} \bigg[1 - \bigg(1+\frac{m_2^2\Phi_{\rm M}^2}{2\Phi_\star^4}\bigg)\frac{m_1^2}{2 \Phi_\star^2}  - \bigg(1 + \frac{2\Phi_{\rm M}}{\Phi_\star}\bigg)\frac{m_2^2}{2\Phi_\star^2}\bigg]\;.
\end{aligned}
\end{equation}
Since the location of the false vacuum must fall into the regime of $V_2$, i.e., $\Phi_{\rm M} < \Phi_2 < 0$, this gives the condition\footnote{Actually, the condition (\ref{eq:con_m1}) automatically leads to $\Phi_2 \leq \Phi_{\rm P} < 0$.}, 
\begin{align}\label{eq:con_m1}
    2 + \frac{m_1^2 \Phi_{\rm M}}{ \Phi_\star^3} < 0 \;.
\end{align}
For illustrative purposes, we plot the potential (\ref{eq:piecewise_potential}) in Fig.~\ref{fig:piecewise_potential}.
In the plot, the red line corresponds to the exponential potential, whereas the blue and the green lines refer to the potentials $V_1$ and $V_2$, respectively. 
The vertical dashed lines indicate the two matching points at $\Phi = 0$ and $\Phi = \Phi_2 = -2.34 \Phi_\star$, from right to left. 
Clearly, the potential (\ref{eq:piecewise_potential}) is unbounded from below for $\Phi \geq 0$.

In each regime of the potential (\ref{eq:piecewise_potential}), the background instanton solution $\bar{\Phi}(\tilde{\rho})$ can be obtained analytically. 
As obtained before in section~\ref{sec:cubic_potential}, in the exponential-potential regime the solution for $\bar{\Phi}(\tilde{\rho})$ is given by  
\begin{align}\label{eq:bg_exp}
    \bar{\Phi}(\tilde{\rho}) = -\Phi_\star\log(\tilde{\rho}) \;; \qquad 0<\tilde{\rho}\leqslant1 \;.
\end{align}
Notice that we disregard the solution regular at $\tilde{\rho} = 0$.\footnote{The solution (\ref{eq:bg_exp}) is uniquely determined when imposing the condition $\Phi'_0 \rightarrow \infty$ at $\tilde{\rho} = 0$. 
See appendix~\ref{app:reverse_process} for detailed discussion.}
In the $V_1$ regime, the solution reads 
\begin{align}\label{eq:sol_phi_V1}
    \bar{\Phi} (\tilde{\rho}) = - \frac{2\Phi_\star^3}{m_1^2} + \frac{c_1}{\tilde{\rho}} J_1\big(\frac{m_1}{\Phi_\star}\tilde{\rho}\big) + \frac{c_2}{\tilde{\rho}}Y_1\big(\frac{m_1}{\Phi_\star} \tilde{\rho}\big) \;; \qquad 1\leqslant\tilde{\rho}\leqslant\tilde{\rho}_2 \,,
\end{align}
where $J_1$ and $Y_1$ are Bessel functions of the first and second kinds, respectively, and $c_1$, $c_2$ are constants.
In the $V_2$ regime, we have 
\begin{align}\label{eq:sol_phi_0_V2}
    \bar{\Phi}(\tilde{\rho}) = \Phi_{\rm M} + \frac{c_3}{\tilde{\rho}} K_1\big(\frac{m_2}{\Phi_\star}\tilde{\rho}\big) + \frac{c_4}{\tilde{\rho}}I_1\big( \frac{m_2}{\Phi_\star}\tilde{\rho}\big) \;; \qquad \tilde{\rho}\geqslant\tilde{\rho}_2 \;,
\end{align}
where $K_1$ and $I_1$ are modified Bessel functions of the first and second kinds respectively, and $c_3$, $c_4$ are constants.
Note that the Bessel functions $K_1(z)$ and $I_1(z)$ behave as $e^{-z}$ and $e^{z}$ as $z \rightarrow \infty$, respectively.
We see that in order for $\bar{\Phi}$ asymptotically goes to $\Phi_{\rm M}$ as $\tilde{\rho} \rightarrow \infty$, we impose $c_4 = 0$.

\begin{figure}[t]
\includegraphics[width=0.45\textwidth]{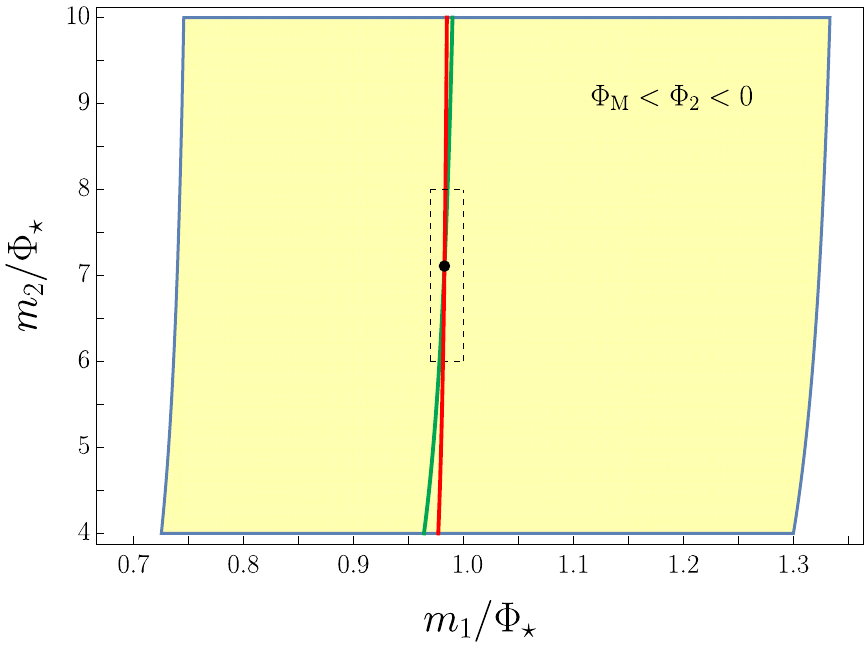}
\ \
\includegraphics[width=0.45\textwidth]{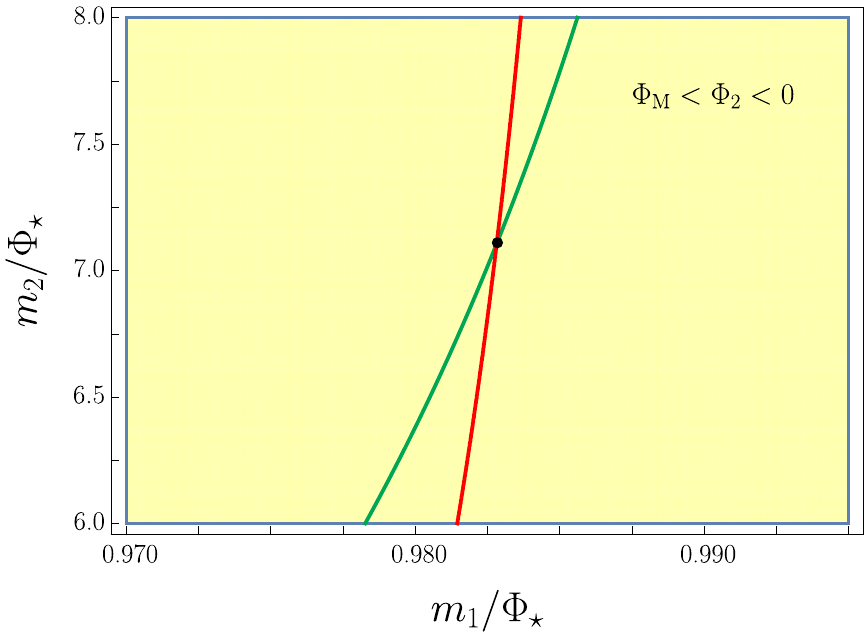}
\caption{~\textit{Left panel}: Parameter space of $m_1/\Phi_\star$ and $m_2/\Phi_\star$.
The condition $\Phi_{\rm M} < \Phi_2 < 0$ [Eq.~\eqref{eq:con_bg}] is satisfied in the yellow region.
The green line refers to the condition \eqref{eq:con_bg_Phi_2}: $\bar{\Phi}(\tilde{\rho}_2) = \Phi_2$, while the red curve corresponds to the condition $c_8 = 0$.
We choose $\tilde{\rho}_2 = 5$ and $\alpha = 0.5$.
The black dot is $\{0.98,7.11\}$, for which there exist both background solution $\bar{\Phi}(\tilde{\rho})$ and the regular deformation.
The corresponding values of $\Phi_2$ and $\Phi_{\rm M}$ are respectively $\Phi_2 = -2.34\Phi_\star$ and $\Phi_{\rm M} = -2.35\Phi_\star$.  
The small dashed box indicates a region near the intersection point, with a zoomed-in view displayed in the {right panel}.
\textit{Right panel}: The figure zooms in the neighborhood of the intersection point at which the conditions \eqref{eq:con_bg}, \eqref{eq:con_bg_Phi_2} and $c_8=0$ are satisfied. This numerically shows the existence of $O(4)$-symmetric background solution $\bar\Phi(\rho)$ given in \eqref{eq:bg_exp}--\eqref{eq:sol_phi_0_V2} and the radial component of the deformation $f_L(\rho)$ which breaks $O(4)$-symmetry, given in \eqref{eq:sol_f_expo}--\eqref{eq:sol_second_f_L}.} 
\label{fig:contour_plot_piecewise} 
\end{figure}

Then, we use the continuity conditions of $\bar{\Phi}(\tilde{\rho})$ and its first derivative at $\tilde{\rho} = 1$ ($\bar{\Phi} = 0$), so that the constants $c_1$ and $c_2$ are fixed in terms of $m_1$. We have 
\begin{align}\label{eq:app:c1_c2}
    c_1 = \frac{\pi \Phi_\star}{2} Y_1\big(\frac{m_1}{\Phi_\star}\big) - \frac{\pi \Phi_\star^2}{m_1} Y_{2} \big(\frac{m_1}{\Phi_\star}\big)  \;, \qquad
    c_2 = -\frac{\pi\Phi_\star}{2} J_1\big(\frac{m_1}{\Phi_\star}\big)  + \frac{\pi\Phi_\star^2}{m_1} J_{2} \big(\frac{m_1}{\Phi_\star}\big)  \;.
\end{align}
For convenience, we introduce 
\begin{align}
    H_1 &\equiv 2m_1 \Phi_\star J_0\big(\frac{m_1}{\Phi_\star}\big) + (m_1^2 - 4\Phi_\star^2)J_1\big(\frac{m_1}{\Phi_\star}\big) \;, \\
    H_2 &\equiv 2m_1 \Phi_\star Y_0\big(\frac{m_1}{\Phi_\star}\big) + (m_1^2 - 4\Phi_\star^2) Y_1\big(\frac{m_1}{\Phi_\star}\big) \;.
\end{align}
With the above definitions of $H_1$ and $H_2$, the continuity of both $\bar{\Phi}$ and $\bar{\Phi}'$ at $\tilde{\rho} = \tilde{\rho}_2$ yields
\begin{align}
    c_3 &= - \frac{m_2\tilde{\rho}_2^2\Phi_\star^2}{m_1^2}\bigg(2 + \frac{m_1^2 \Phi_{\rm M}}{\Phi_\star^3}\bigg)I_2\big(\frac{m_2\tilde{\rho}_2}{\Phi_\star}\big) + \frac{\pi H_1 \tilde{\rho}_2}{2m_1^2} \bigg[m_1 I_1\big(\frac{m_2\tilde{\rho}_2}{\Phi_\star}\big)Y_0\big(\frac{m_1\tilde{\rho}_2}{\Phi_\star}\big) - m_2 I_0\big(\frac{m_2 \tilde{\rho}_2}{\Phi_\star}\big)Y_1\big(\frac{m_1 \tilde{\rho}_2}{\Phi_\star}\big)\bigg] \nonumber \\
    &\hspace{4mm} -\frac{\pi H_2 \tilde{\rho}_2}{2m_1^2} \bigg[m_1 I_1\big(\frac{m_2\tilde{\rho}_2}{\Phi_\star}\big)  J_0\big(\frac{m_1\tilde{\rho}_2}{\Phi_\star}\big) - m_2 I_0\big(\frac{m_2 \tilde{\rho}_2}{\Phi_\star}\big)J_1\big(\frac{m_1 \tilde{\rho}_2}{\Phi_\star}\big)\bigg] \;, \\
    c_4 &= - \frac{m_2\tilde{\rho}_2^2\Phi_\star^2}{m_1^2}\bigg(2 + \frac{m_1^2 \Phi_{\rm M}}{\Phi_\star^3}\bigg)K_2\big(\frac{m_2\tilde{\rho}_2}{\Phi_\star}\big) - \frac{\pi H_1 \tilde{\rho}_2}{2m_1^2}\bigg[m_1 Y_0\big(\frac{m_1\tilde{\rho}_2}{\Phi_\star}\big)K_1\big(\frac{m_2\tilde{\rho}_2}{\Phi_\star}\big) + m_2 Y_1\big(\frac{m_1 \tilde{\rho}_2}{\Phi_\star}\big)K_0\big(\frac{m_2 \tilde{\rho}_2}{\Phi_\star}\big)\bigg] \nonumber \\ 
    &\hspace{4mm} + \frac{\pi H_2 \tilde{\rho}_2}{2m_1^2}\bigg[m_1 J_0\big(\frac{m_1\tilde{\rho}_2}{\Phi_\star}\big)K_1\big(\frac{m_2\tilde{\rho}_2}{\Phi_\star}\big) + m_2 J_1\big(\frac{m_1 \tilde{\rho}_2}{\Phi_\star}\big)K_0\big(\frac{m_2 \tilde{\rho}_2}{\Phi_\star}\big)\bigg]\;.
\end{align}
Then, requiring that $c_4$ vanishes (to exclude the exponentially growing mode) gives the condition for $\Phi_{\rm M}$ in terms of $m_1$, $m_2$ and $\tilde{\rho}_2$.
Therefore, combining the condition $c_4 = 0$ and the inequality (\ref{eq:con_m1}), we obtain 
\begin{align}\label{eq:con_bg}
  m_2  K_0\big(\frac{m_2\tilde{\rho}_2}{\Phi_\star}\big) \bigg[H_2 J_1\big(\frac{m_1\tilde{\rho}_2}{\Phi_\star}\big) - H_1 Y_1\big(\frac{m_1\tilde{\rho}_2}{\Phi_\star}\big) \bigg] + m_1K_1\big(\frac{m_2\tilde{\rho}_2}{\Phi_\star}\big)\bigg[H_2 J_0\big(\frac{m_1\tilde{\rho}_2}{\Phi_\star}\big) - H_1 Y_0\big(\frac{m_1\tilde{\rho}_2}{\Phi_\star}\big) \bigg]  \leq 0 \;.
\end{align}
In addition to (\ref{eq:con_bg}), we identify the solution \eqref{eq:sol_phi_0_V2} at $\tilde{\rho} = \tilde{\rho}_2$ with $\Phi_2$ in (\ref{paracon}), i.e.,
\begin{align}\label{eq:con_bg_Phi_2}
\bar{\Phi}(\tilde{\rho}_2) = \Phi_2 \,.
\end{align}
The conditions (\ref{eq:con_bg}) and (\ref{eq:con_bg_Phi_2}) must be satisfied at $\tilde{\rho}_2 > 1$, so that $\bar{\Phi}(\tilde{\rho})$ can be smoothly connected and asymptotically goes to the false vacuum $\Phi_{\rm M}$. 
In Fig.~\ref{fig:contour_plot_piecewise}, we demonstrate the existence of such solutions by the green line in the yellow region.
The yellow region represents the region satisfying the condition (\ref{eq:con_bg}), and the green curve refers to the condition (\ref{eq:con_bg_Phi_2}). 
After fixing $\tilde{\rho}_2$ and imposing the above condition, the remaining parameters are $m_1$ and $\alpha$. Note that $\alpha$ is completely arbitrary. 
Thus, for each value of $\alpha$, there exists a one-parameter family of the potential that allows the existence of singular instantons with finite action. 

Let us now analyze the dynamics of the small deformation $\delta\Phi(\rho, \vec{\theta}\,)$ in \eqref{eq:ansatz}.
As discussed before in section~\ref{sec:cubic_potential}, we introduced the new variable $f_L \equiv \tilde{\rho}^{3/2} A_L/\Phi_\star$ whose equation of motion is given by Eq.~(\ref{eq:EoM_fluctuation}).
Note that similar to the background solutions we can obtain analytic solutions for $f_L(\tilde{\rho})$ in each regime of the potential (\ref{eq:piecewise_potential}).
In the exponential-potential regime, we have
\begin{align}\label{eq:sol_f_expo}
    f_L(\tilde{\rho}) = a_1 \tilde{\rho}^{\frac{1}{2} + \sqrt{j}} \;; \qquad 0<\tilde{\rho}\leqslant1\,.
\end{align}
Recall that $j \equiv (L - 1)(L + 3)$.
In the $V_1$ regime, we obtain
\begin{align}\label{eq:sol_first_f_L}
    f_L(\tilde{\rho}) =& \sqrt{\tilde{\rho}}\left[c_5 J_{L+1}\big(\frac{m_1}{\alpha^2\Phi_\star}\tilde{\rho}\big) +  c_6 Y_{L+1}\big(\frac{m_1}{\alpha^2\Phi_\star}\tilde{\rho}\big) \right] \;; \qquad 1\leqslant\tilde{\rho}\leqslant\tilde{\rho}_2\;,
\end{align}
where $c_5$ and $c_6$ are constants. 
Similarly, the solution in the $V_2$ regime reads 
\begin{align}\label{eq:sol_second_f_L}
    f_L(\tilde{\rho}) = &\sqrt{\tilde{\rho}}\left[c_7 K_{L+1}\big(\frac{m_2}{\alpha^2\Phi_\star}\tilde{\rho}\big) +  c_8 I_{L+1}\big(\frac{m_2}{\alpha^2\Phi_\star}\tilde{\rho}\big) \right] \;; \qquad \tilde{\rho}\geqslant\tilde{\rho}_2 \,,
\end{align}
where $c_7$ and $c_8$ are constants. 
Then imposing the continuity of $f_L(\tilde{\rho})$ and its first derivative at $\tilde{\rho} = 1$, we obtain
\begin{align}
    c_5 &= \frac{\pi}{2} \bigg[\frac{m_1}{\alpha^2 \Phi_\star}Y_L\big(\frac{m_1}{\alpha^2 \Phi_\star}\big) -(L+1 + \sqrt{j}) Y_{L+1}\big(\frac{m_1}{\alpha^2 \Phi_\star}\big)\bigg]a_1 \;, \\
    c_6 &= -\frac{\pi}{2} \bigg[\frac{m_1}{\alpha^2 \Phi_\star}J_L\big(\frac{m_1}{\alpha^2\Phi_\star}\big) - (L+1 + \sqrt{j}) J_{L+1}\big(\frac{m_1}{\alpha^2\Phi_\star}\big)\bigg]a_1 \;,
\end{align}
Furthermore, we use the matching conditions at $\tilde{\rho} = \tilde{\rho}_2$ to determine $c_7$ and $c_8$ in terms of $\alpha$, $m_1$, $m_2$, $\tilde{\rho}_2$ and $a_1$. We now have
\begin{align}
    c_7 &=  \frac{\pi \tilde{\rho}_2}{2\alpha^4 \sin(L\pi)}  \bigg[ \frac{m_1}{\Phi_\star} \mathcal{D}^+_{-L,-1-L} J_L\big(\frac{m_1\tilde{\rho}_2}{\alpha^2\Phi_\star}\big) I_{1+L}(\frac{m_2\tilde{\rho}_2}{\alpha^2\Phi_\star})  - \frac{m_1}{\Phi_\star}\mathcal{D}^-_{L,1+L}J_{-L}\big(\frac{m_1\tilde{\rho}_2}{\alpha^2\Phi_\star}\big)I_{1+L}\big(\frac{m_2\tilde{\rho}_2}{\alpha^2\Phi_\star}\big)  \nonumber \\ 
    & \hspace{5mm} -  \frac{ m_2 }{\Phi_\star} \mathcal{D}^-_{L,1+L}J_{-1-L}\big(\frac{m_1\tilde{\rho}_2}{\alpha^2\Phi_\star}\big)I_L\big(\frac{m_2\tilde{\rho}_2}{\alpha^2\Phi_\star}\big)  -\frac{ m_2 }{\Phi_\star} \mathcal{D}^+_{-L,-1-L}J_{1+L}\big(\frac{m_1\tilde{\rho}_2}{\alpha^2\Phi_\star}\big)I_L\big(\frac{m_2\tilde{\rho}_2}{\alpha^2\Phi_\star}\big) \bigg]a_1 \;, \\ 
    c_8 &= \frac{\pi \tilde{\rho}_2 }{2\alpha^4}\bigg\{\bigg[\frac{m_1}{\Phi_\star} J_L\big(\frac{m_1\tilde{\rho}_2}{\alpha^2\Phi_\star} \big) K_{1+L}(\frac{m_2\tilde{\rho}_2}{\alpha^2\Phi_\star}) + \frac{m_2}{\Phi_\star} J_{1+L}\big(\frac{m_1\tilde{\rho}_2}{\alpha^2\Phi_\star}\big) K_L\big(\frac{m_2\tilde{\rho}_2}{\alpha^2\Phi_\star}\big)\bigg]\bigg[\frac{m_1}{\Phi_\star} Y_L\big(\frac{m_1}{\alpha^2\Phi_\star}\big) - \alpha^2\tilde{j} Y_{1+L}\big(\frac{m_1}{\alpha^2\Phi_\star}\big)\bigg] \nonumber \\
    &\hspace{5mm} - \bigg[\frac{m_1}{\Phi_\star} Y_L\big(\frac{m_1\tilde{\rho}_2}{\alpha^2\Phi_\star}\big) K_{1+L}\big(\frac{m_2\tilde{\rho}_2}{\alpha^2\Phi_\star}\big) + \frac{m_2}{\Phi_\star} Y_{1+L}\big(\frac{m_1\tilde{\rho}_2}{\alpha^2\Phi_\star} \big) K_L\big(\frac{m_2\tilde{\rho}_2}{\alpha^2\Phi_\star}\big)\bigg]\mathcal{D}^-_{L,1+L}\bigg\}a_1 \;. \label{eq:c_8}\end{align}
where $\tilde{j} \equiv 1+L+\sqrt{j}$ and we have defined
\begin{align}
    \mathcal{D}^{\pm}_{A,B} \equiv \frac{m_1}{\Phi_\star} J_{A}\big(\frac{m_1}{\alpha^2\Phi_\star}\big) \pm \alpha^2\tilde{j} J_{B}\big(\frac{m_1}{\alpha^2\Phi_\star}\big) \;,
\end{align}
with the indices $A$, $B$ referring to $\{-1-L, -L, L, 1+L\}$.
Finally, we require the solution (\ref{eq:sol_second_f_L}) to be regular as $\tilde{\rho} \rightarrow\infty$, this fixes $c_8(m_1, m_2, \tilde{\rho}_2, a_1, \alpha) = 0$. 
For given values of $\tilde{\rho}_2$ and $\alpha$, this condition provides a relation between $m_1$ and $m_2$. Thus for a fixed $\alpha$, there is a one-parameter family of regular solutions.

\begin{figure}[t]
\includegraphics[width=0.435\textwidth]{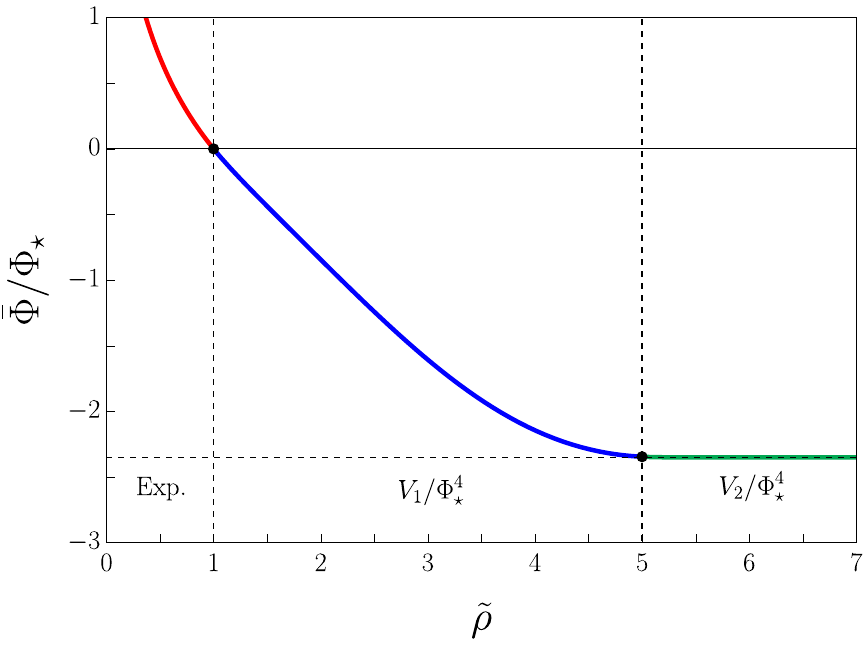}
\ \
\includegraphics[width=0.45\textwidth]{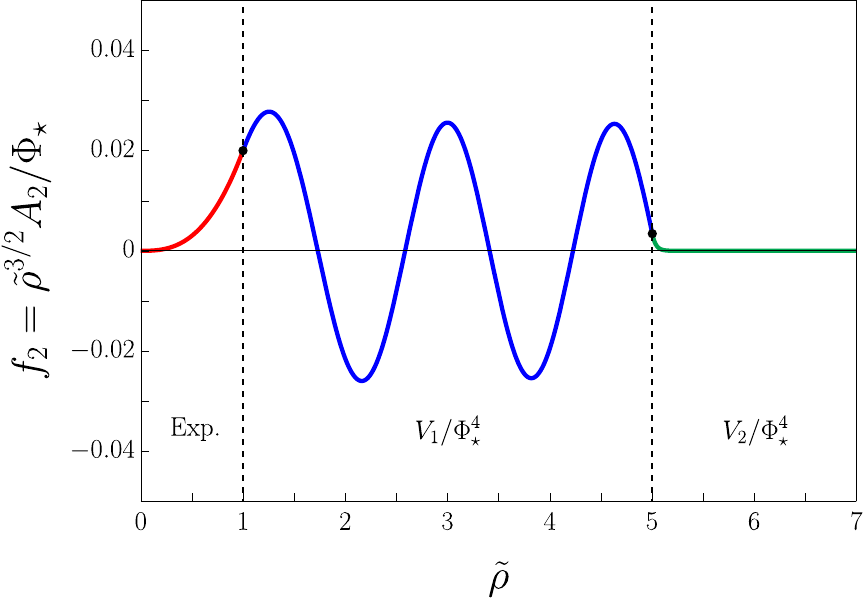}
\caption{~The solutions for $\bar{\Phi}(\tilde{\rho})$ (\textit{left panel}) and $f_2(\tilde{\rho})$ with $L=2$ (\textit{right panel}). 
The solutions in the exponential potential, $V_1$ and $V_2$ regimes are respectively represented by the red, blue and green colors.
We set $\alpha = 0.5$, $m_1 = 0.98\Phi_\star$ and $m_2 = 7.11\Phi_\star$, so that $\Phi_2 = -2.34\Phi_\star$ and $\Phi_{\rm M} = -2.35\Phi_\star$.
In the right panel we appropriately rescale the solution $f_2(\tilde{\rho})$ with the free parameter $a_1$.
The horizontal dashed line denotes the location for the false vacuum ($\Phi_{\rm M} = -2.35\Phi_\star$).
The vertical dashed lines in both panels refer to the two matching points at $\tilde{\rho} = 1$ and $\tilde{\rho}_2 = 5$, from left to right.}
\label{fig:fluctuation_piecewise} 
\end{figure}

Since both the background and the deformation have a one-parameter family of solutions, 
there exists a perturbatively non-$O(4)$-symmetric singular instanton with finite action if the two one-parameter families intersect with each other in the parameter space.  
In Fig.~\ref{fig:contour_plot_piecewise}, we show the existence of both $\bar{\Phi}$ and regular deformations by the intersection point where the conditions [$c_8 = 0$, Eqs.~(\ref{eq:con_bg}) and (\ref{eq:con_bg_Phi_2})] are satisfied.
The figure in the right panel only zooms in the neighborhood of the intersection point at which all of the conditions mentioned above are satisfied.
In both panels the regular condition $c_8 = 0$ with $\tilde{\rho}_2 = 5$ and $\alpha = 0.5$ is represented by the red curve, together with the conditions (\ref{eq:con_bg}) (the yellow region) and $\bar{\Phi}(\tilde{\rho}) = \Phi_2$ (green curve).
More precisely, at the intersection point $\{m_1/\Phi_\star = 0.98, m_2/\Phi_\star = 7.11\}$ all of the conditions [$c_8 = 0$, $\bar{\Phi}(\tilde{\rho}) = \Phi_2$ and Eq.~(\ref{eq:con_bg})] are simultaneously satisfied.
In other words, for particular values of $\tilde{\rho}_2$ and $\alpha$ there exists a point where both the background $\bar{\Phi}$ and the regular deformation can be realized.
Thus we have explicitly shown the existence of regular deformed instanton solutions with finite action in this example.
For different values of $\tilde{\rho}_2$ and $\alpha$, one is required to check the existence of the regular deformed solutions case by case. 

In Fig.~\ref{fig:fluctuation_piecewise} we plot both the background solution $\bar{\Phi}(\tilde{\rho})$ (left panel) and the deformation $f_2(\tilde{\rho})$ with $L=2$ (right panel) as a function of $\tilde{\rho}$. 
Note that in the right panel we appropriately rescale $f_2(\tilde{\rho})$ with the free parameter $a_1$.
The solutions in the exponential potential, $V_1$ and $V_2$ regimes are labeled by red, blue and green respectively.
The vertical dashed lines denote the two matching points at $\tilde{\rho} = 1$ and $\tilde{\rho} = 5$, from left to right.
We explicitly see that $\bar{\Phi}(\tilde{\rho})$ diverges as $\tilde{\rho} \to 0$, while it goes to $\Phi_{\rm M}$ (false vacuum) as $\tilde{\rho} \to \infty$.
Moreover, the deformation solution $f_2(\tilde{\rho})$ is regular everywhere, especially at the origin and infinity.
Therefore, in this model there exist regular and small deformations around the singular instanton. 
We note that the solution we found has four nodes, indicating that there exist four additional negative modes. 
Fortunately, the fact that $i^4 = 1$ implies that the computation of the decay rate will not be affected by the existence of these negative modes.
Nevertheless, it may signal an instability of the solution. 
This suggests that there may indeed exist a finitely deformed instanton with lower action.

It is interesting to point out that in the limit $\alpha \ll 1$ we find multiple one-parameter families of the potentials that allow the existence of both a singular O(4)-symmetric solution with finite action and regular small deformations with vanishing second-order action. 
Specifically, one can check that in this limit of $\alpha$ the condition $c_8 = 0$ can be realized when $m_1 = m_2 = m$ and 
\begin{align}\label{eq:n_m}
   \frac{m \tilde{\rho}_2}{\alpha^2 \Phi_\star} - \frac{L\pi}{2} = n \pi\;, 
\end{align}
where $n$ is an integer. Using the above choice together with the conditions for the existence of a background solution, Eqs.~(\ref{eq:con_bg}) and (\ref{eq:con_bg_Phi_2}), we find there exist multiple one-parameter families that allow deformed instanton solutions.
Notice that in the limit $\alpha \gg 1$, the coefficient $c_8$ is dominated by a single term which is positive definite; therefore, it is impossible to satisfy the condition $c_8 = 0$.

For completeness, we compute the bounce action evaluated on the solution we found, i.e., the deformed instanton solution at the intersection point in Fig.~\ref{fig:contour_plot_piecewise}. 
First, we recall that in the presence of deformations the bounce action up to second order in perturbations is given by $B[\Phi] = B_0[\bar{\Phi}] + \epsilon^2 B_2[\delta\Phi]$, see Eq.~(\ref{eq:action_Phi1}). 
For regular deformations, $B_2 = 0$ as we have seen in section~\ref{sec:bounda_singular}.
Therefore, we may focus on the zeroth-order action $B_0[\bar{\Phi}]$,
\begin{align}\label{eq:pw_bounce}
    B_0[\bar{\Phi}] = \frac{2 \pi^2}{\alpha^4} \int_0^{\tilde{\rho}_{\rm f}}\mathrm{d}\tilde{\rho}~\tilde{\rho}^3  \bigg[\frac{1}{2\Phi_\star^2}\bigg(\frac{\mathrm{d}\bar{\Phi}}{\mathrm{d}\tilde{\rho}}\bigg)^2 + \frac{\tilde{V}(\bar{\Phi}) - \tilde{V}(\Phi_{\rm FV})}{\Phi_\star^4}\bigg] \;,
\end{align}
where we have defined $\tilde{V} \equiv V/\alpha^4$. 
It is now evident that the dimensionless parameter $\alpha$ does not appear in the integrand of Eq.~(\ref{eq:pw_bounce}).
We then plug the solution shown in Fig.~\ref{fig:fluctuation_piecewise} into the action and perform the numerical integration in the range $\tilde{\rho} \in [0,\tilde{\rho}_{\rm f}]$ with $\tilde{\rho}_{\rm f} = 10$.
We obtain
\begin{align}\label{eq:pw_bounce_num}
    B_0[\bar{\Phi}] = \mathcal{C}\cdot\frac{2 \pi^2}{\alpha^4}  \;,
\end{align}
where $\mathcal{C}$ is a numerical factor. 
For the solution existing on the intersection point of Fig.~\ref{fig:contour_plot_piecewise}, we have $\mathcal{C} \simeq 7.41$.\footnote{This value corresponds to $B_0 \sim \mathcal{O}(100)$ for $\alpha = 1$, which is in the validity of the WKB approximation. In fact, the semiclassical approximation is expected to break down when $\alpha \gtrsim \mathcal{O}(10)$, implying that quantum fluctuations become important.}
Eq.~(\ref{eq:pw_bounce_num}) is the bounce action for the singular instantons with regular deformations. 
We see that as expected the action $B_0[\bar{\Phi}]$ scales as $\alpha^{-4}$ with numerical factor depending on the model parameters.
We note that, for $\alpha \ll 1$, the action is much larger than unity, which justifies the use of semiclassical approximation. 
Notice that we computed $B_0[\bar{\Phi}]$ only on a single solution we found on the parameter space of $m_1$ and $m_2$, shown in Fig.~\ref{fig:contour_plot_piecewise}. 
For other values of $m_1$ and $m_2$, numerical integration must be done on a case-by-case basis.

To conclude this section, we emphasize again that a small and regular deformation can be regarded as a zero mode on the $O(4)$-symmetric singular solutions.
Our finding in this case gives rise to the possibility that there may exist non-$O(4)$-symmetric solutions with finite action with its action lower than the $O(4)$-symmetric instanton, once we allow singular solutions with finite action.

\section{Conclusions}\label{sec:conclusion}
We have studied a possibility of extending Coleman's proof in \cite{Coleman:1977th} to the case where the $O(4)$-symmetric instanton solution exhibits a singular behavior at the true-vacuum bubble ($\rho \to 0$). 
Specifically, we focused on the case where the potential is unbounded from below and the solution does not obey Coleman's boundary condition (\ref{eq:con_coleman}).
This indeed violates one of the assumptions of Coleman's theorem, allowing the possibility of non-Coleman type instantons with finite action.

In section~\ref{sec:gen_potential}, we have derived a general form of the instanton potential with the assumptions that the value of the field as well as its first derivative blows up as $\rho \rightarrow 0$, but its corresponding action is finite. 
We note that for the potential that allows the singular instanton with finite action, there exists no regular instanton solutions because the boundary condition at false vacuum selects a unique solution.
In section~\ref{sec:cuspy_sol}, 
we have discussed the case when a cusp in the field configuration appears at the origin and identified the shape of the potential that allows such a cuspy solution. However, we have not performed a detailed analysis as our main focus is on singular solutions.
In section~\ref{sec:small_deform}, we have studied small anisotropic deformations around the $O(4)$-symmetric instanton, and clarified that physically meaningful  deformations are of quadrupole or higher multipole nature.  
In section~\ref{sec:bounda_singular}, we have expressed the action up to second order in the deformation parameter $\epsilon$. We have found the condition for the action to be finite, and found that, for regular deformations, the second-order action is solely determined in terms of their boundary behavior, see Eq.~(\ref{eq:s_2_phi_1}).

In section~\ref{sec:con_example}, we have analyzed two concrete examples: 
The cubic potential and the piecewise quadratic potential on the false vacuum side, while the potential on the true vacuum side is exponential for both cases.
The two examples differ by the detailed potential on the negative-$\Phi$ side, whereas for $\Phi \geq 0$ both have an unbounded exponential potential. 
For the cubic-potential case (section~\ref{sec:cubic_potential}), we have obtained the solutions for $\bar{\Phi}$ and numerically computed the bounce action as a function of parameter $b$. 
However, in this example both regular and singular deformations around the $O(4)$-symmetric instanton solutions with finite action do not exist.  
In section~\ref{sec:piecewise_qud}, we have studied the piecewise quadratic case, whose result is concisely presented in \cite{Sasaki:2024vul}. 
In this case, we have found that there exist solutions for both the background $\bar{\Phi}$ and the regular deformations with finite action for a one-parameter family of the model parameters.
Interestingly, these regular deformations do not make any contribution to the total bounce action at quadratic order in perturbation. 
Namely, the bounce action evaluated on the regular deformed instanton solutions is the same as the $O(4)$-symmetric instanton solutions. 
Note that, as briefly discussed above Eq.~(\ref{eq:n_m}), the existence of additional four negative modes signals an instability of the instanton, which implies that there may exist a finitely deformed solution whose action is lower than that of the $O(4)$-symmetric instanton.

There are several future directions we would like to study. 
First, it would be interesting to generalize our result in this paper to a gravitational system (see \cite{Cohn:1998et,Gratton:1999ya,Dunne:2006bt,Ai:2023yce,Oshita:2023pwr,Oshita:2021aux} for related work). 
Second, it would be worth extending our analysis to the case of multi-field models, see e.g.~\cite{Sugimura:2011tk,Masoumi:2017trx}. 
Third, since the second-order action vanishes on the regular deformations, it is worth investigating the action at higher orders. 
Related to this direction, it would be nice if we could study finite anisotropic deformations non-perturbatively. 
This is definitely an interesting issue for future study.

\section*{Acknowledgements}
We thank J.~R.~Espinosa, S.~Mukohyama, N.~Oshita, R.~Saito, K.~Takahashi and M.~Yamaguchi for useful discussions. 
M.S. and V.Y. are supported by World Premier International Research Center Initiative (WPI Initiative), MEXT, Japan. 
V.Y. is supported by grants for development of new faculty staff, Ratchadaphiseksomphot Fund, Chulalongkorn University and by the NSRF via the Program Management Unit for Human Resources \& Institutional Development, Research and Innovation Grant No.\ B39G680009.
Y.Z. is supported by  the Fundamental Research Funds for the Central Universities, and by the Project 12475060 and 12047503 supported by NSFC, Project 24ZR1472400 sponsored by Natural
Science Foundation of Shanghai, and Shanghai Pujiang Program 24PJA134.
This work is also supported in part by JSPS KAKENHI No. 24K00624.
\appendix
\section{Spherical harmonics in $(n-1)$ dimensions}\label{app:spherical_harmonics}
In this appendix we review some properties of the $(n-1)$-dimensional spherical harmonics. See \cite{doi:10.1063/1.527513} for more details. The $(n-1)$-dimensional spherical harmonics $Y_L^M(\vec{\theta})$ satisfy the following eigenvalue equation:
\begin{align}
    \Delta_{S^{n-1}} Y_L^M(\vec{\theta}) = -L(L + n - 2) Y^M_L(\vec{\theta}) \;,
\end{align}
where $\vec{\theta} = (\theta_1, \theta_2, \ldots, \theta_{n-1})$ and $M$ is a multi-index characterizing the magnetic quantum numbers satisfying $|m_1| \leq m_2 \leq \cdots \leq m_{n-2} \leq L \equiv m_{n-1}$. Note that the operator $\Delta_{S^{n-1}}$ is the Laplace operator defined by 
\begin{align}
    \Delta_{S^{n - 1}} \equiv \sin^{2 - n}\theta_{n-1} \frac{\partial}{\partial \theta_{n-1}}\bigg(\sin^{n - 2} \theta_{n-1} \frac{\partial}{\partial \theta_{n-1}}\bigg) + \sin^{-2}\theta_{n-1} \Delta_{S^{n - 2}} \;.
\end{align}
Here we choose $\theta_{n-1}$ to be an axial coordinate in a spherical coordinate system on $S^{n - 1}$. Note that in our case where $n = 4$ the coordinate $\theta_3$ corresponds to our coordinate $\theta$ introduced in (\ref{eq:flat_metric}). Actually, written in terms of other known functions, the functions $Y_L^M(\vec{\theta})$ are
\begin{align}\label{eq:n_spherical_H}
    Y_L^{m_1\cdots m_{n-1}}(\vec{\theta}) = \frac{1}{\sqrt{2\pi}} e^{im_1\theta_1} \prod_{i = 2}^{n-1} {}_{i}\bar{P}_{m_i}^{m_{i-1}}(\theta_i) \;,
\end{align}
where 
\begin{equation}\label{eq:asso_Legen}
\begin{aligned}
    {}_{k}\bar{P}_{i}^{j}(\theta) = \ &\sqrt{\frac{(2i + k -1)(i + j + k - 2)!}{2(i - j)!}}  \sin^{\frac{2-k}{2}}(\theta)P_{\frac{2i + k - 2}{2}}^{-\frac{2j + k - 2}{2}}(\cos\theta) \;, \\
    P_b^{-a}(x) = \ &\frac{1}{\Gamma(1 + a)} \bigg(\frac{1-x}{1 + x}\bigg)^{a/2} {}_2F_1\bigg(-b, b+1; 1+a; \frac{1 - x}{2}\bigg) \;, 
\end{aligned}
\end{equation}
for $|1 - x| < 2$. The function $P_b^{-a}$ are the associated Legendre functions and ${}_2F_1$ is the hypergeometric function.
As usual, these functions form a complete set which satisfies the following properties on the $(n-1)$-sphere:
\begin{equation}
\begin{aligned}
    \int {\rm d}\Omega_{S^{n-1}} Y^M_L(\vec{\theta})^* Y_{L'}^{M'}(\vec{\theta}) &= \delta_{LL'} \delta^{MM'} \;, \label{eq:orth_Y}\\
    \sum_{L,M} Y^M_L(\vec{\theta})^* Y_L^M(\vec{\theta}') &= \frac{1}{\sqrt{\gamma}} \delta^{(n-1)}(\vec{\theta} - \vec{\theta}') \;,
\end{aligned}
\end{equation}
where ${\rm d}\Omega_{S^{n-1}}$ denotes the volume element on the sphere $S^{n-1}$ and $\gamma$ is a normalization factor.

\section{Regular solution with sextic form}\label{app:sextic_solution}
In this appendix, we are going to regularize our singular instanton solution, $\bar{\Phi}_{\rm sing}(\tilde{\rho}) = -\Phi_\star \log(\tilde{\rho})$, which is an exact solution in the exponential-potential regime, see section~\ref{sec:con_example}. 
Here, for simplicity, we regularize such a solution by connecting it to the following ansatz: 
\begin{align}\label{eq:sol_reg_phi_sextic}
    \bar{\Phi}_{\rm reg}(\tilde{\rho}) = \Phi_0 - \frac{\lambda_2}{2} \tilde{\rho}^2 - \frac{\lambda_4}{4}\tilde{\rho}^4 - \frac{\lambda_6}{6}\tilde{\rho}^6  \;,
\end{align}
where $\Phi_0$ is the location where the scalar field stops at $\tilde{\rho} = 0$, and $\lambda_2, \lambda_4$ and $\lambda_6$ are constants with mass dimension one. 
Note that once the field stops evolving at $\bar{\Phi}(\tilde{\rho} = 0) = \Phi_0$ one can always connect it with another potential which possesses a minimum.  
It is worth noting that alternative approaches to regularizing our singular solution may exist; however, such methods may require the use of numerical methods.

Then, using the ansatz (\ref{eq:sol_reg_phi_sextic}) in Eq.~(\ref{eq:EOM_phi_zeroth}) it is straightforward to derive the potential,
\begin{align}\label{eq:potential_sextic}
   \frac{ V_{\rm reg}(\tilde{\rho})}{\alpha^4} = \Lambda_0 + 2 \Phi_\star^2 \tilde{\rho}^2 \bigg[\lambda_2^2  + \frac{5}{4} \lambda_4 \lambda_2 \tilde{\rho}^2 + \frac{1}{2}(\lambda_4^2 + 2 \lambda_2 \lambda_6)\tilde{\rho}^4  + \frac{7}{8}\lambda_4 \lambda_6  \tilde{\rho}^6 + \frac{2}{5}\lambda_6^2 \tilde{\rho}^{8} \bigg] \;,
\end{align}
where $\Lambda_0$ is a constant.
In fact, one can straightforwardly solve for $\tilde{\rho}^2$ in terms of $\Phi$ using Eq.~(\ref{eq:sol_reg_phi_sextic}), giving
\begin{align}\label{eq:root_rho2}
    \tilde{\rho}^2 = -\frac{\lambda_4}{2\lambda_6} + \frac{\mathcal{H}^{1/3}}{2\lambda_6} + \frac{\lambda_4^2 - 4\lambda_2 \lambda_6}{2\lambda_6 \mathcal{H}^{1/3}} \;, 
\end{align}
where we have defined
\begin{align}
    \mathcal{H} \equiv -\lambda_4^3 + 6 \lambda_2 \lambda_4 \lambda_6 + 24 \lambda_6^2(\Phi_0 - \Phi) \;,
\end{align}
and we have disregarded other complex roots.
Thus, using Eq.~(\ref{eq:root_rho2}) in (\ref{eq:sol_reg_phi_sextic}) one obtains the potential as a function of $\Phi$, but here we omit the full expression $V_{\rm reg}(\Phi)$ due to its length.

Let us now consider the matching conditions at $\tilde{\rho} = \tilde{\rho}_1$ and $\bar{\Phi} = \Phi_1$. 
Note before that, in this case, at first we have seven parameters: $\{\Phi_0, \lambda_2, \lambda_4, \lambda_6, \Lambda_0, \tilde{\rho}_1, \Phi_1\}$.
Then, by imposing the continuity conditions of the potential at $\tilde{\rho} = \tilde{\rho}_1$ we obtain
\begin{equation}\label{eq:match_Lambda_0_lambda_3}
\begin{aligned}
    \Lambda_0 &= -\frac{\Phi_\star^2}{\tilde{\rho}_1^2} \left[\Phi_\star^2 + 2 \bar{\lambda}_2^2 + \frac{5}{2}\bar{\lambda}_2 \bar{\lambda}_4 + \bar{\lambda}_4^2 + 2 \bar{\lambda}_2 \bar{\lambda}_6  + \frac{35}{20}\bar{\lambda}_4 \bar{\lambda}_6 
    + \frac{8}{10}\bar{\lambda}_6^2 \right] \;, \\
    \bar{\lambda}_2 &= -\frac{5\bar{\lambda}_4}{4} - \frac{3\bar{\lambda}_6}{2}  + \frac{1}{4}\sqrt{8\Phi_\star^2 + (\bar{\lambda}_4 + 2 \bar{\lambda}_6)^2} \;.
\end{aligned}
\end{equation}
where we have, for convenience, normalized the parameters $\lambda_i$'s as $\lambda_2 = \bar{\lambda}_2/\tilde{\rho}_1^2$, $\lambda_4 = \bar{\lambda}_4/\tilde{\rho}_1^4$ and $\lambda_6 = \bar{\lambda}_6/\tilde{\rho}_1^6$.
Moreover, matching the singular solution $\bar{\Phi}_{\rm sing}(\tilde{\rho})$ with the regular solution (\ref{eq:sol_reg_phi_sextic}) at $\tilde{\rho} = \tilde{\rho}_1$ yields 
\begin{align}\label{eq:match_lambda_1_Phi_f}
    \bar{\lambda}_6 = - \frac{1}{2} (\Phi_\star + \bar{\lambda}_4 ) \;, \qquad 
    \Phi_0 = -\Phi_\star \log(\tilde{\rho}_1) + \frac{2}{3}\Phi_\star - \frac{\bar{\lambda}_4}{12}  \;.
\end{align}
We see that now using Eq.~(\ref{eq:match_lambda_1_Phi_f}) in (\ref{eq:match_Lambda_0_lambda_3}) we obtain
\begin{align}\label{eq:Lambda_0_lam3}
    \Lambda_0 = -\frac{\Phi_\star^4}{\tilde{\rho}_1^2} \left[\frac{21}{5} - \frac{29\bar{\lambda}_4 }{40 \Phi_\star} + \frac{3\bar{\lambda}_4^2}{40\Phi_\star^2}\right] \;, \qquad 
    \bar{\lambda}_2 = \frac{1}{2}(3\Phi_\star - \bar{\lambda}_4) \;.
\end{align}
Therefore, the parameters $\{\Phi_0, \bar{\lambda}_2, \bar{\lambda}_6, \Lambda_0\}$ can be expressed in terms of $\tilde{\rho}_1$ and $\bar{\lambda}_4$. 
It is important to note that $\Phi_1$ must be fixed to be $\Phi_1 = -\Phi_\star \log(\tilde{\rho}_1)$, as it is required by the singular solution at $\tilde{\rho} = \tilde{\rho}_1$. 
Hence, there are only two free parameters and, for convenience, we choose them to be $\{\bar{\lambda}_4, \tilde{\rho}_1\}$.
In fact, from Eq.~(\ref{eq:match_lambda_1_Phi_f}) we can rewrite it as 
\begin{align}
    \Phi_0 - \Phi_1 = \frac{2}{3}\bigg(\Phi_\star - \frac{\bar{\lambda}_4}{8}\bigg)  \;,
\end{align}
which must be positive, so that our regular solution is monotonic in the range $0 \leq \tilde{\rho} \leq \tilde{\rho}_1$.
Thus, demanding that $\Phi_0 > \Phi_1$ results in the allowed region in the parameter between $\bar{\lambda}_4$ and $\tilde{\rho}_1$ where $\bar{\lambda}_4/\Phi_\star < 8$. 
Also, we note that when $\tilde{\rho}_1 \to 0$ one has $\Phi_1 \to \infty$ ($\Phi_0 \to \infty$), which corresponds to the limit where we recover our singular instanton solution $\bar{\Phi}_{\rm sing}(\tilde{\rho})$. 
More explicitly, using Eqs.~(\ref{eq:match_lambda_1_Phi_f}) and (\ref{eq:Lambda_0_lam3}) in (\ref{eq:sol_reg_phi_sextic}) we thus obtain
\begin{align}\label{eq:reg_sol_final}
    \bar{\Phi}_{\rm reg}(\tilde{\rho}) = -\Phi_\star \log(\tilde{\rho}_1) + \frac{(\tilde{\rho}^2 - \tilde{\rho}_1^2)}{12\tilde{\rho}_1^6} \bigg[\Phi_\star (\tilde{\rho}^4 + \tilde{\rho}^2 \tilde{\rho}_1^2 - 8 \tilde{\rho}_1^4) + \bar{\lambda}_4  (\tilde{\rho}^2 - \tilde{\rho}_1^2)^2\bigg] \;. 
\end{align}
As expected, in the form written above, it is evident that $\bar{\Phi}_{\rm reg}(\tilde{\rho}_1) = -\Phi_\star \log(\tilde{\rho}_1)$.

\begin{figure}[t]
\includegraphics[width=0.55\textwidth]{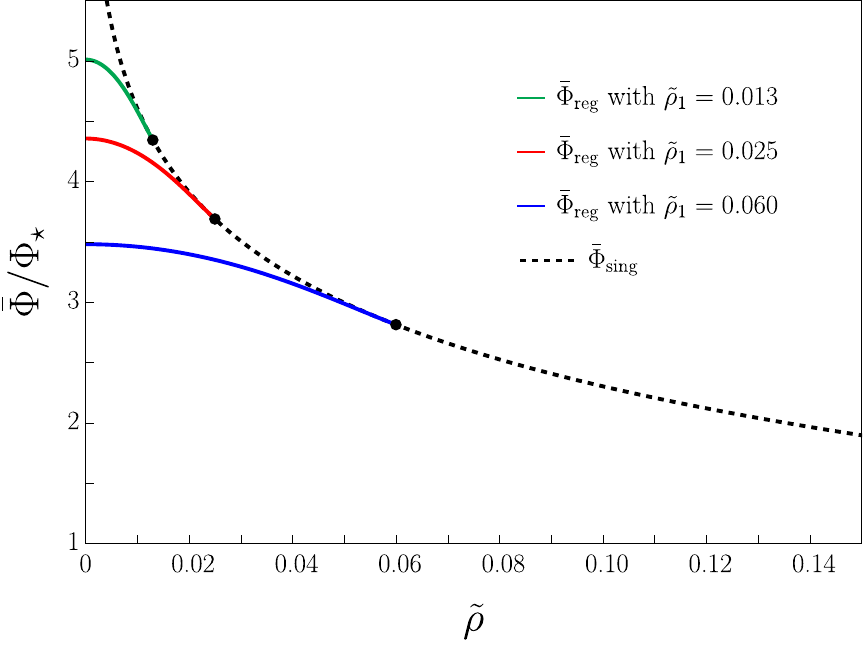}
\caption{~The solution $\bar{\Phi}/\Phi_\star$ as a function of $\tilde{\rho}$. The green, red and blue solid lines represent the regular solutions with $\tilde{\rho}_1 = 0.013$, $0.030$ and $0.06$, respectively. The black dashed line shows the singular solution, $\bar{\Phi}(\tilde{\rho}) = -\Phi_\star\log(\tilde{\rho})$.
The matching location for the cases where $\tilde{\rho}_1 = 0.013$, $0.025$ and $0.06$ is represented by the black dot.
Here we set $\bar{\lambda}_4 = 0$.}
\label{fig:sol_sextic} 
\end{figure}

\begin{figure}[t]
\includegraphics[width=0.6\textwidth]{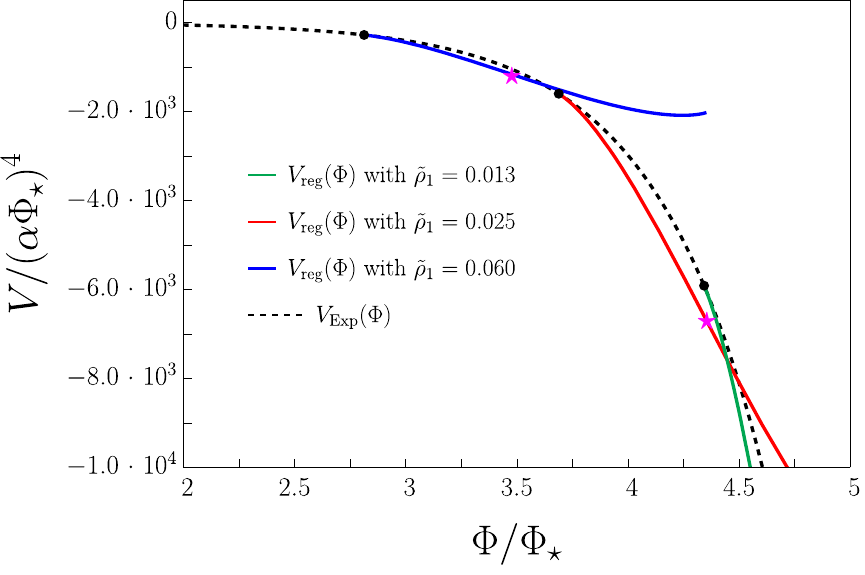}
\caption{~The regularized and the exponential potentials as a function of $\Phi$. The green, red and blue solid lines represent the regularized potentials with $\tilde{\rho}_1 = 0.013$, $0.025$ and $0.06$, respectively. The black dashed line shows the exponential potential, $V_{\rm Exp}(\Phi) = -\alpha^4 \Phi_\star^4 \exp(2\Phi/\Phi_\star)$.
The matching location for the cases where $\tilde{\rho}_1 = 0.013$, $0.025$ and $0.06$ is represented by the black dot.
We do not show the point $\Phi_0$ for the green line as it is located outside the range we are plotting. 
The points $\Phi_0 = 3.48 \Phi_\star$ for the blue line and $\Phi_0 = 4.36\Phi_\star$ for the red line are indicated by the magenta star. 
Here we set $\bar{\lambda}_4 = 0$.}
\label{fig:pol_sextic} 
\end{figure}

Actually, the solution (\ref{eq:reg_sol_final}) can be simply realized when setting $\bar{\lambda}_4 = 0$. 
In this case, we have 
\begin{align}\label{eq:reg_sol_exp}
    \bar{\Phi}_{\rm reg}(\tilde{\rho}) = -\Phi_\star \log(\tilde{\rho}_1) + \frac{\Phi_\star}{12\tilde{\rho}_1^6} \big(\tilde{\rho}^2 - \tilde{\rho}_1^2\big)\big(\tilde{\rho}^4 + \tilde{\rho}^2 \tilde{\rho}_1^2 - 8 \tilde{\rho}_1^4\big) \;.
\end{align}
Using the above expression, in Fig.~\ref{fig:sol_sextic} we plot the regular solutions with $\tilde{\rho}_1 = 0.013$ (green solid line), $\tilde{\rho}_1 = 0.025$ (red solid line) and $\tilde{\rho}_1 = 0.060$ (blue solid line), all of which are smoothly connected with the singular solution (black dashed line).
Clearly, we see that as $\tilde{\rho} \to 0$ the solution $\bar{\Phi}(\tilde{\rho})$ is finite and approaches $\Phi_0$. 
For illustrative purposes, in Fig.~\ref{fig:pol_sextic} we plot the potential $V_{\rm reg}(\Phi)$ in the range $\Phi_1 \leq \Phi$ with $\tilde{\rho}_1 = 0.013$ (green line), $\tilde{\rho}_1 =  0.025$ (red line) and $\tilde{\rho}_1 = 0.06$ (blue line), all of which are connected with the exponential potential $V_{\rm Exp}(\Phi)$ at $\Phi = \Phi_1$. 
It is important to note that in the plot for the potential one has to stop at $\Phi(\tilde{\rho} = 0) = \Phi_0$, as shown by the magenta star. 
Moreover, it is useful to point out that although the solution stops at $\Phi_0$, the potential (\ref{eq:potential_sextic}) possesses a minimum, which is given by the condition:
\begin{align}\label{eq:con_true_sextic}
    \bar{\lambda}_1^2 + 3 \bar{\lambda}_2 \bar{\lambda}_6 \bigg(\frac{\tilde{\rho}_{\rm T}}{\tilde{\rho}_1}\bigg)^4 + 2 \bar{\lambda}_6^2 \bigg(\frac{\tilde{\rho}_{\rm T}}{\tilde{\rho}_1}\bigg)^8 = 0 \;,
\end{align}
where we have set $\bar{\lambda}_4 = 0$ and $\tilde{\rho}_{\rm T}$ denotes the location at the minimum (true vacuum).
Note that the condition (\ref{eq:con_true_sextic}) was derived from taking a derivative of the potential (\ref{eq:potential_sextic}) wrt. $\tilde{\rho}^2$ and setting it to zero. 
In fact, Eq.~(\ref{eq:con_true_sextic}) can be analytically solved for $\tilde{\rho}_{\rm T}$. 
After that, it is straightforward to obtain $\Phi_{\rm T}$ (true-vacuum location) using Eq.~(\ref{eq:root_rho2}) and $\tilde{\rho}_{\rm T}$. 
Here we omit the expression of $\Phi_{\rm T}$ since we do not need it. 
Note that, as mentioned before, it is also possible to connect our potential (\ref{eq:potential_sextic}) to other potentials which have a local minimum at $\Phi = \Phi_0$, so that the true vacuum $\Phi_{\rm T}$ can be realized.

For completeness, let us now compute the action evaluated on the regular solution (\ref{eq:sol_reg_phi_sextic}). 
Note that here we omit the computation of the action for $\bar{\Phi} < 0$, as it was already given in the main text. 
Let us define the action difference $\Delta S$ as
\begin{align}
    \Delta S &\equiv S_0^{(\rm reg)}[\bar{\Phi}] - S_0^{(\rm sing)}[\bar{\Phi}] \\ 
   &= \frac{2\pi^2}{\alpha^4} \int_0^{\tilde{\rho}_1} \de \tilde{\rho}\, \tilde{\rho}^3 \left\{\bigg[\frac{1}{2\Phi_\star^2}\bigg(\frac{\de \bar{\Phi}_{\rm reg}}{\de \tilde{\rho}}\bigg)^2 + \frac{V_{\rm reg}(\bar{\Phi}_{\rm reg})}{(\alpha\Phi_\star)^4}\bigg] 
   - \bigg[\frac{1}{2\Phi_\star^2}\bigg(\frac{\de \bar{\Phi}_{\rm sing}}{\de \tilde{\rho}}\bigg)^2 + \frac{V_{\rm Exp}(\bar{\Phi}_{\rm sing})}{(\alpha\Phi_\star)^4}\bigg] \right\}\;, \label{eq:action_sextic}
\end{align}
where $S_0^{(\rm reg)}[\bar{\Phi}]$ is the action evaluated on the regular instanton and $S_0^{(\rm sing)}[\bar{\Phi}]$ is the singular-instanton action in the range $0 < \tilde{\rho} < \tilde{\rho}_1$. 
Using Eqs.~(\ref{eq:reg_sol_final}) and (\ref{eq:potential_sextic}) in (\ref{eq:action_sextic}) we therefore obtain
\begin{align}\label{eq:action_diff_sextic}
    \Delta S = -\frac{9 \pi^2 \tilde{\rho}_1^2}{70 \alpha^4} + \frac{\bar{\lambda}_4 \pi^2 \tilde{\rho}_1^2 (\bar{\lambda}_4  - 19 \Phi_\star) }{1680\alpha^4 \Phi_\star^2} \;,
\end{align}
where we have used $\Phi_1 = -\Phi_\star \log(\tilde{\rho}_1)$, $\bar{\Phi}_{\rm sing}(\tilde{\rho}) = -\Phi_\star \log(\tilde{\rho})$ and $V_{\rm Exp}(\bar{\Phi}_{\rm sing}) = -\alpha^4 \Phi_\star^4 \exp(2\bar{\Phi}_{\rm sing}/\Phi_\star)$.
We see that this action difference is finite as $\tilde{\rho}_1 \to 0$. 
In addition, the condition $\bar{\lambda}_4/\Phi_\star < 8$, which guarantees $\Phi_0 > \Phi_1$, tells us that the second term on the RHS of Eq.~(\ref{eq:action_diff_sextic}) is always negative.  
This implies that in this case $\Delta S$ is always negative, i.e., $S_0^{(\rm sing)}[\bar{\Phi}] > S_0^{(\rm reg)}[\bar{\Phi}]$.
Note that whether this is generically the case for regularized instanton solutions remains an open question for future investigation.
\section{Uniqueness and model-parameter dependence}\label{app:reverse_process}
In this appendix, we consider the reverse process of matching the solutions in section~\ref{sec:piecewise_qud}, i.e., we start from the solution that asymptotically goes to the false vacuum $\Phi_{\rm M}$, and match it with other solutions in the regimes of $V_1(\Phi)$ and $V_{\rm Exp}(\Phi)$ respectively, to show the uniqueness of the singular solution with finite action for a specific choice of parameters.
We also show that a small deviation in model parameters from this specific choice leads to a singular solution with singular action.
As we will see, this shows that the singular behavior as $\tilde{\rho} \to 0$, $\bar{\Phi}(\tilde{\rho}) = -\Phi_\star\log(\tilde{\rho})$, crucially depends on the model parameters $m_1$ and $m_2$ one is choosing. 

Let us recall the solution in the $V_2$ regime [Eq.~(\ref{eq:sol_phi_0_V2})],
\begin{align}\label{eq:sol_phi_0_V2_app}
    \bar{\Phi}(\tilde{\rho}) = \Phi_{\rm M} + \frac{c_3}{\tilde{\rho}} K_1\big(\frac{m_2}{\Phi_\star}\tilde{\rho}\big) \;; \qquad \tilde{\rho} \geq \tilde{\rho}_2
\end{align}
where we have set $c_4 = 0$ since, as explained above, we now start from the solution that goes to $\Phi_{\rm M}$ as $\tilde{\rho} \to \infty$.
Then, matching the above solution with the solution (\ref{eq:sol_phi_V1}) at $\tilde{\rho} = \tilde{\rho}_2$ we have 
\begin{equation}\label{eq:reverse_c1_c2}
\begin{aligned}
    c_1 &= \frac{c_3 m_2}{m_1 \mathcal{F}_1} K_2\big(\frac{m_2 \tilde{\rho}_2}{\Phi_\star}\big)Y_1\big(\frac{m_1 \tilde{\rho}_2}{\Phi_\star}\big) - \frac{1}{\mathcal{F}_1}\bigg[\Phi_{\rm M}\tilde{\rho}_2 + \frac{2\tilde{\rho}_2\Phi_\star^3}{m_1^2} + c_3 K_1\big(\frac{m_2 \tilde{\rho}_2}{\Phi_\star}\big) \bigg]Y_2\big(\frac{m_1 \tilde{\rho}_2}{\Phi_\star}\big) \;,  \\ 
    c_2 &=  -\frac{c_3 m_2 }{m_1 \mathcal{F}_1} K_2\big(\frac{m_2 \tilde{\rho}_2}{\Phi_\star}\big) J_1\big(\frac{m_1 \tilde{\rho}_2}{\Phi_\star}\big) + \frac{1}{\mathcal{F}_1}\bigg[\Phi_{\rm M}\tilde{\rho}_2 + \frac{2\tilde{\rho}_2\Phi_\star^3}{m_1^2} + c_3 K_1\big(\frac{m_2 \tilde{\rho}_2}{\Phi_\star}\big) \bigg]\;, 
\end{aligned}
\end{equation}
where we have defined 
\begin{align}
    \mathcal{F}_1 \equiv J_2 \big(\frac{m_1 \tilde{\rho}_2}{\Phi_\star}\big) Y_1 \big(\frac{m_1 \tilde{\rho}_2}{\Phi_\star}\big) - J_1 \big(\frac{m_1 \tilde{\rho}_2}{\Phi_\star}\big) Y_2 \big(\frac{m_1 \tilde{\rho}_2}{\Phi_\star}\big) \;.
\end{align}
Also, using the condition that $\bar{\Phi}(\tilde{\rho}_2) = \Phi_2$ yields 
\begin{align}\label{eq:revere_Phi_M}
    \Phi_{\rm M} = - \frac{2}{m_1^2} - \frac{c_3}{\tilde{\rho}_2}\bigg(1 + \frac{m_2^2}{m_1^2}\bigg) K_1\big(\frac{m_2 \tilde{\rho}_2}{\Phi_\star}\big) \;,
\end{align}
where we have used Eq.~(\ref{paracon}) for the expression of $\Phi_2$.
We now see that using Eqs.~(\ref{eq:reverse_c1_c2}) and (\ref{eq:revere_Phi_M}) the solution (\ref{eq:sol_phi_V1}) in the $V_1$ regime can be solely determined in terms of $m_1$, $m_2$, $\tilde{\rho}_2$ and $c_3$. 
In principle, one can numerically solve the equation of motion (\ref{eq:EOM_phi_zeroth}) for $\bar{\Phi}(\tilde{\rho})$ in the exponential-potential regime given the initial conditions at $\tilde{\rho} = 1$ from the solution (\ref{eq:sol_phi_V1}). 
By doing so, one can clearly see how varying the model parameters $m_1$ and $m_2$ (with $\tilde{\rho}_2$ and $c_3$ held fixed) changes the behavior of $\bar{\Phi}(\tilde{\rho})$ in the range $0 < \tilde{\rho} \leq 1$ (exponential-potential regime).

Since the EOM of $\bar{\Phi}(\tilde{\rho})$ in the exponential-potential regime is non-linear, it is useful to analytically obtain the conditions of the remaining model parameters, under which the solution $\bar{\Phi}(\tilde{\rho}) = -\Phi_\star\log(\tilde{\rho})$ can be realized for $0 < \tilde{\rho} \leq 1$.
To do so, we now match the solution and its derivative at $\tilde{\rho} = 1$. 
The continuity of $\bar{\Phi}(\tilde{\rho})$ at $\tilde{\rho} = 1$ gives
\begin{align}
    c_3 &= -\frac{4\Phi_\star^4}{m_1 m_2 \pi \tilde{\rho}_2}\bigg[m_2 J_2\big(\frac{m_1 \tilde{\rho}_2}{\Phi_\star}\big) K_1\big(\frac{m_2 \tilde{\rho}_2}{\Phi_\star}\big) Y_1\big(\frac{m_1 }{\Phi_\star}\big)
    + m_1 J_1\big(\frac{m_1 \tilde{\rho}_2}{\Phi_\star}\big) K_2\big(\frac{m_2 \tilde{\rho}_2}{\Phi_\star}\big) Y_1 \big(\frac{m_1 }{\Phi_\star}\big) \nonumber \\
    &\hspace{4mm} - m_1  J_1 \big(\frac{m_1 }{\Phi_\star}\big) K_2\big(\frac{m_2 \tilde{\rho}_2}{\Phi_\star}\big) Y_1 \big(\frac{m_1 \tilde{\rho}_2}{\Phi_\star}\big)
    - m_2 J_1 \big(\frac{m_1 }{\Phi_\star}\big) K_1\big(\frac{m_2 \tilde{\rho}_2}{\Phi_\star}\big) Y_2\big(\frac{m_1 \tilde{\rho}_2}{\Phi_\star}\big)\bigg]^{-1} \;.
\end{align}
Then, using the above expression for $c_3$ in the continuity condition for $\bar{\Phi}'(\tilde{\rho})$ at $\tilde{\rho} = 1$ yields
\begin{align}\label{eq:con_app_m1m2}
  \mathcal{F}_2 \left[ J_0 \big(\frac{m_1\tilde{\rho}_2}{\Phi_\star}\big) K_1\big(\frac{m_2\tilde{\rho}_2}{\Phi_\star}\big)
  - \frac{\mathcal{F}_4}{m_1} J_1 \big(\frac{m_1\tilde{\rho}_2}{\Phi_\star}\big)  \right]
  - \mathcal{F}_3\left[ K_1\big(\frac{m_2\tilde{\rho}_2}{\Phi_\star}\big) Y_0\big(\frac{m_1\tilde{\rho}_2}{\Phi_\star}\big)  
  - \frac{\mathcal{F}_3\mathcal{F}_4}{m_1} Y_1(\frac{m_2\tilde{\rho}_2}{\Phi_\star})\right] = 0 \;,
\end{align}
where we have defined
\begin{align}
    \mathcal{F}_2 &\equiv 2m_1 \Phi_\star Y_0 \big(\frac{m_1}{\Phi_\star}\big) + (m_1^2 - 4\Phi_\star^2 )Y_1\big(\frac{m_1}{\Phi_\star}\big) \;, \\
    \mathcal{F}_3 &\equiv 2m_1 \Phi_\star J_0 \big(\frac{m_1}{\Phi_\star}\big) + (m_1^2 - 4\Phi_\star^2 )J_1\big(\frac{m_1}{\Phi_\star}\big) \;, \\
    \mathcal{F}_4 &\equiv \frac{m_1^2}{m_2} K_0\big(\frac{m_2 \tilde{\rho}_2}{\Phi_\star}\big) + \frac{2\Phi_\star}{\tilde{\rho}_2}\bigg(1 + \frac{m_1^2}{m_2^2}\bigg) K_1 \big(\frac{m_2 \tilde{\rho}_2}{\Phi_\star}\big) \;.
\end{align}
We see that for a fixed value of $\tilde{\rho}_2$, the condition (\ref{eq:con_app_m1m2}) gives a line, similar to the green line in Fig.~\ref{fig:contour_plot_piecewise}, in the space of $m_1$ and $m_2$. 
Furthermore, we impose the inequality $\Phi_{\rm M} < \Phi_2 < 0$, which yields 
\begin{align}
    K_1\big(\frac{m_2 \tilde{\rho}_2}{\Phi_\star}\big) &\bigg[m_2 J_2\big(\frac{m_1 \tilde{\rho}_2}{\Phi_\star}\big) K_1\big(\frac{m_2 \tilde{\rho}_2}{\Phi_\star}\big) Y_1\big(\frac{m_1 }{\Phi_\star}\big)
    + m_1 J_1\big(\frac{m_1 \tilde{\rho}_2}{\Phi_\star}\big) K_2\big(\frac{m_2 \tilde{\rho}_2}{\Phi_\star}\big) Y_1 \big(\frac{m_1 }{\Phi_\star}\big) \nonumber \\
    &- m_1  J_1 \big(\frac{m_1 }{\Phi_\star}\big) K_2\big(\frac{m_2 \tilde{\rho}_2}{\Phi_\star}\big) Y_1 \big(\frac{m_1 \tilde{\rho}_2}{\Phi_\star}\big)
    - m_2 J_1 \big(\frac{m_1 }{\Phi_\star}\big) K_1\big(\frac{m_2 \tilde{\rho}_2}{\Phi_\star}\big) Y_2\big(\frac{m_1 \tilde{\rho}_2}{\Phi_\star}\big)\bigg]^{-1} < 0 \;. \label{eq:app_ineq}
\end{align}
The inequality above leads to the allowed region in the parameter space of $m_1$ and $m_2$, similar to the yellow region in Fig.~\ref{fig:contour_plot_piecewise}. 
We checked that there exist the allowed region where both conditions (\ref{eq:con_app_m1m2}) and (\ref{eq:app_ineq}) are satisfied.

Suppose that we do not fix the solution in the exponential-potential regime to be $\bar{\Phi}(\tilde{\rho}) = -\Phi_\star\log(\tilde{\rho})$ and we also want to avoid solving the non-linear differential equation numerically. 
In this situation, it is useful to consider a small variation of the model parameters $m_1$ and $m_2$: $m_1 \to m_1 + \delta m_1$ and $m_2 \to m_2 + \delta m_2$, where $\delta m_1$ and $\delta m_2$ are small changes of $m_1$ and $m_2$. 
In this case, the solution (\ref{eq:sol_phi_V1}) in the range $1 \leq \tilde{\rho} \leq \tilde{\rho}_2$ becomes 
\begin{align}\label{eq:sol_Phi_V1_app}
    \bar{\Phi} (\tilde{\rho}) = - \frac{2\Phi_\star^3}{(m_1 + \delta m_1)^2} + \frac{c_1}{\tilde{\rho}} J_1\big(\frac{(m_1 + \delta m_1 )}{\Phi_\star}\tilde{\rho}\big) + \frac{c_2}{\tilde{\rho}}Y_1\big(\frac{(m_1 + \delta m_1)}{\Phi_\star} \tilde{\rho}\big) \;. 
\end{align}
Note that since the coefficients $c_1$ and $c_2$ depend on $m_1$ and $m_2$ [Eq.~(\ref{eq:reverse_c1_c2})], the changes of $m_1$ and $m_2$ also affect $c_1$ and $c_2$ accordingly.
Thus, this implies that having these small changes affects the conditions at $\bar{\Phi} = 0$.
In other words, these deviations lead to $\bar{\Phi} = 0 + h_1$ and $\bar{\Phi}' = -1 + h_2$ at $\tilde{\rho} = \tilde{\rho}_0 \equiv 1 + \Delta \tilde{\rho}$ with the condition that $h_1 = 0$ at $\tilde{\rho} = \tilde{\rho}_0$, where $h_1$ and $h_2$ are given by Eq.~(\ref{eq:sol_Phi_V1_app}) and its first derivative expanded at first order in $\delta m_1$ and $\delta m_2$.
In this case, these small changes of the initial conditions at $\tilde{\rho} = \tilde{\rho}_0$ induce a small deviation from the logarithmic solution in the exponential-potential regime,
\begin{align}
\bar{\Phi}(\tilde{\rho}) = \bar{\Phi}_0(\tilde{\rho}) + \Delta \bar{\Phi}(\tilde{\rho}) \;; \qquad 
\bar{\Phi}_0(\tilde{\rho}) = -\Phi_\star \log\bigg(\frac{\tilde{\rho}}{\tilde{\rho}_0}\bigg) \;,
\end{align}
where $\Delta \bar{\Phi}(\tilde{\rho})$ is the deviation generated by small variations of $m_1$ and $m_2$. 
Notice that $\Delta \bar{\Phi}(\tilde{\rho})$ only depends on $\tilde{\rho}$ since we are still at the level of background $O(4)$-symmetric solution. 
Similar to the analysis in section~\ref{sec:small_deform}, the EOM of $\Delta \bar{\Phi}(\tilde{\rho})$ reads 
\begin{align}
   \frac{\de^2\Delta \bar{\Phi}}{\de \tilde{\rho}^2} + \frac{3}{\tilde{\rho}} \frac{\de \Delta \bar{\Phi}}{\de \tilde{\rho}} - \frac{1}{\alpha^4 \Phi_\star^2}\frac{\de^2 V}{\de\Phi^2}\bigg|_{\bar{\Phi}_0} \Delta \bar{\Phi} = 0  \;.
\end{align}
Note that the above equation corresponds to Eq.~(\ref{eq:A(rho)}) for $L = 0$.
For notational simplicity, we redefine $\tilde{\rho}$ by $\tilde{\rho}/\tilde{\rho}_0$ in the following.
Using $V_{\rm Exp}(\Phi) = -\alpha^4 \Phi_\star^4 \exp(2\Phi/\Phi_\star)$, we thus obtain
\begin{align}\label{eq:sol_Delta_bar_Phi}
    \Delta \bar{\Phi}(\tilde{\rho}) = \frac{q_1}{\tilde{\rho}} \cos\big(\sqrt{3}\,\log(\tilde{\rho})\big) + \frac{q_2}{\tilde{\rho}} \sin\big(\sqrt{3}\,\log(\tilde{\rho})\big) \;,
\end{align}
where $q_1$ and $q_2$ are constants.
We note that unless we vary $m_1$ and $m_2$ along the direction indicated by the green line in Fig.~6, $q_1$ and $q_2$ are generally non-vanishing. 
We see that the solution (\ref{eq:sol_Delta_bar_Phi}) has a peculiar oscillation from the $\cos\big(\sqrt{3}\,\log(\tilde{\rho})\big)$ and $\sin\big(\sqrt{3}\,\log(\tilde{\rho})\big)$ functions; however, it diverges as $\tilde{\rho} \to 0$. 
This implies that any deviations from the solution $\bar{\Phi}_0(\tilde{\rho}) = -\Phi_\star\log(\tilde{\rho})$ result in unstable solutions. 
Moreover, it is straightforward to show that the action evaluated on the solution (\ref{eq:sol_Delta_bar_Phi}) contains divergences as $\tilde{\rho} \to 0$. 
Therefore, this analysis tells us that our singular solution in the exponential-potential regime is unique under the condition that the action must be finite. 
	{}
\bibliographystyle{utphys}
\bibliography{bib_v4.bib}

\end{document}